\documentclass[11pt,a4paper]{article}
\pdfoutput=1
\usepackage{jheppub}

\usepackage{graphicx}

\usepackage{amssymb}
\usepackage{amsmath}
\usepackage{mathrsfs}
\usepackage{dsfont}

\usepackage{tikz}
\usetikzlibrary{trees}
\usetikzlibrary{decorations.pathmorphing}
\usetikzlibrary{decorations.markings}
\usetikzlibrary{calc}

\usepackage{color}

\usepackage{vmargin}
\setmargins  {2.5cm}{1.5cm}
             {15cm}{22.5cm}
             {28pt}{15pt}
             {0pt}{40pt}

\newcommand{\Eq}[1]{Eq.~(\ref{#1})}

\begin{document}

\title{Leading hadronic contributions to the running of the electroweak coupling constants from lattice QCD}


\author[a]{Florian Burger,}
\author[b]{Karl Jansen,}
\author[c]{Marcus Petschlies,}
\author[d]{Grit Pientka}


\affiliation[a]{OakLabs GmbH, Neuendorfstr. 20B, D-16761 Hennigsdorf, Germany}
\affiliation[b]{NIC, DESY, Platanenallee 6, D-15738 Zeuthen, Germany }
\affiliation[c]{Institut f\"ur Strahlen- und Kernphysik, Rheinische Friedrich-Wilhelms-Universit\"at Bonn,
Nussallee 14-16, D-53115 Bonn, Germany}
\affiliation[d]{Humboldt-Universit\"at zu Berlin, Institut f\"ur Physik, Newtonstr. 15, \par D-12489 Berlin, Germany }

\emailAdd{burger@oak-labs.com, karl.jansen@desy.de, marcus.petschlies@hiskp.uni-bonn.de, grit.hotzel@physik.hu-berlin.de}

\abstract{
The quark-connected leading-order hadronic contributions to the running of the electromagnetic fine structure constant, $\alpha_{\rm QED}$, and the
weak mixing angle, $\theta_W$, are determined by a four-flavour lattice QCD computation with twisted mass fermions. Full agreement of the
results with a phenomenological analysis is 
observed with an even comparable statistical uncertainty. We show that  
the uncertainty of the lattice calculation is dominated by systematic effects which 
then leads to significantly larger errors than obtained by the phenomenological analysis.
}

%

\keywords{quantum chromodynamics, lattice QCD, fine structure constant, weak mixing angle, hadronic vacuum polarization}

\arxivnumber{1505.03283}

\maketitle


\section{Introduction}
Finding hints for new physics beyond the standard model (SM) has been a major
objective of particle physics over the past decades. A very promising strategy to detect such effects are 
high precision experimental measurements which are matched by 
equally precise theoretical
predictions. An important ingredient for the precision attainable in a theoretical calculation 
is the knowledge of the coupling constants since they enter the quantum loop 
corrections. 

In this article, we investigate the leading-order hadronic contributions for
two of these couplings, the electromagnetic fine structure constant, 
$\alpha_{\rm QED}$, and the $SU(2)_L$ coupling constant, $\alpha_2$, both related
by the weak mixing angle, $\theta_W$. 
An accurate knowledge of these hadronic contributions is mandatory to  
accomplish sufficiently precise predictions for future high-energy colliders~\cite{Jegerlehner:2011mw} 
or low energy experiments \cite{Hewett:2012ns}.

However,  
the hadronic contributions to the running of $\alpha_{\rm QED}$ 
turn out to be only poorly known 
at the scale of the Z-boson mass.
Compared to $\alpha_{\rm QED}$ at zero momentum transfer, there is 
a five orders of magnitude
loss of precision when $\alpha_{\rm QED}$ is taken at the Z-scale 
turning
$\alpha_{\rm QED}(M_Z^2)$  
into one of the least determined 
input parameters of the standard model~\cite{Hagiwara:2011af}.

Phenomenologically, the leading hadronic
contribution to the running of 
$\alpha_{\rm QED}$ originating from hadronic vacuum 
polarisation effects, $\Delta \alpha_{\rm QED}^{\rm hvp}$, 
is determined 
from a dispersion relation and experimental $e^+e^-$ scattering data for the hadronic
cross-sections~\cite{Jegerlehner:2008rs, Jegerlehner:2011mw, Hagiwara:2011af}. 
Although new data has
recently become available,
the present analysis does not lead to a sufficient improvement of the error 
which would be needed for the requirements 
of future collider experiments~\cite{Hagiwara:2011af}.

In principle, lattice QCD calculations would be an ideal tool to determine
the hadronic contributions to electroweak observables such as $\alpha_{\rm QED}$
or $\alpha_2$ considered here. However, presently the precision that can be obtained
from such lattice QCD computations is usually still lower than from the  
phenomenological analyses.  
Nevertheless, the steady progress which is taking place in lattice QCD calculations 
promises to make it an 
expedient alternative to the phenomenological results in the future.
In fact, as we will demonstrate here, even with our present simulations 
the statistical uncertainty already matches the phenomenological error of $\Delta \alpha_{\rm QED}^{\rm hvp}$ and $\Delta^{\rm hvp} \sin^2 \theta_W$. 

$\Delta \alpha_{\rm QED}^{\rm hvp}$ has first been investigated on the lattice  for two
dynamical twisted mass fermions~\cite{Renner:2012fa}. Preliminary results 
incorporating also dynamical strange and charm quarks 
for one selected momentum value have been reported in~\cite{Feng:2012gh}. 
Another determination of $\Delta \alpha_{\rm QED}^{\rm hvp}$, following 
the approach suggested in~\cite{Jegerlehner:2008rs} has been performed 
in~\cite{Francis:2014yga}.

Here, we present our results obtained on the
$N_f=2+1+1$ ensembles of
the European twisted mass collaboration~\cite{Baron:2010bv, Baron:2010th}. We will include 
an estimate of the systematic uncertainties originating from the continuum limit
and from the extrapolation to the physical point for energies ranging from $0$ to $10\,{\rm GeV}^2$.
 
In contrast to $\Delta \alpha_{\rm QED}^{\rm hvp}$, the hadronic 
contributions to the running of the weak mixing angle, $\theta_W$, have not been
studied on the lattice so far. Such a calculation is important since  
the phenomenological determination at low energies cannot only be based on data 
but also needs some assumptions such 
as a partial flavour separation of the cross-section data~\cite{Jegerlehner:1985gq}. 

Lattice calculations can, in contrast, 
provide a first-principle evaluation of the weak mixing angle in the low-momentum
region, where several measurements exist~\cite{Wood:1997zq, Anthony:2005pm, Wang:2014bba, Armstrong:2014tna}.            
In addition, 
due to the great potential of such low energy experiments for unveiling the nature of physics beyond the SM,  
there are also newly planned experimental facilities~\cite{Benesch:2014bas, Chen:2014psa, Becker:2013fya}, 
see also~\cite{Kumar:2013yoa} for a discussion on such experiments. 
Here, 
we present the first lattice QCD calculation of the leading hadronic contribution to the 
weak mixing angle, $\Delta^{\rm hvp} \sin^2 \theta_W$.

\section{The fine structure constant $\alpha_{\rm QED}$}
\label{sec:alpha}
Radiative corrections lead to charge renormalisation 
and thus to the running of the fine structure
constant obtained by summing the one-particle irreducible bubble 
insertions in the photon propagator~\cite{Jegerlehner:1985gq}
\begin{equation}
 \alpha_\mathrm{QED}(Q^2) = \frac{\alpha_0}{1-\Delta\alpha_\mathrm{QED}(Q^2) }\; \mathrm{.}
\label{running}
\end{equation}
Here, $\alpha_0$ is the value at vanishing momentum transfer 
$Q^2=0$, $\alpha_0^{-1}=137.035999173(35)$~\cite{Aoyama:2012wj}. The leading-order
hadronic contribution is given by~\cite{Jegerlehner:2011mw}
\begin{equation}
\Delta \alpha_\mathrm{QED}^{\mathrm{hvp}}(Q^2) = -4 \pi \alpha_0 \Pi_{\mathrm{R}}\left(Q^2\right)
\label{eq:def_alpha}
\end{equation}
and is thus proportional to the subtracted vacuum polarisation function
\begin{equation}
\Pi_{\mathrm{R}}(Q^2)= \Pi(Q^2)- \Pi(0)\; .
\label{eq:Pi_R}
\end{equation}
As mentioned in the introduction, this is 
usually~\cite{Jegerlehner:1985gq,Aoyama:2012wj,Jegerlehner:2011mw} determined by a phenomenological
approach relying on the
once-subtracted dispersion relation~\cite{Jegerlehner:2009ry} which for Euclidean momenta $Q^2$ reads
\begin{equation}
 \Pi_{\mathrm{R}}(Q^2)= \frac{\alpha_0}{3 \pi}\, Q^2\, \int_0^\infty ds \frac{R_{\rm had}(s)}{s(s+Q^2)}
\label{eq:disp}
\end{equation}
and experimental cross-section data for 
\begin{equation}
 R_{\rm had}(s) = \frac{\sigma(e^+e⁻ \rightarrow {\rm hadrons})}{\frac{4\pi \alpha_{\rm QED}^2(s)}{3s}} \, .
\end{equation}

Lattice QCD represents an ab-initio alternative for the calculation of $\Pi_{\mathrm{R}}(Q^2)$, 
since the hadronic vacuum polarisation tensor can be obtained directly in
Euclidean space-time from the correlator of two electromagnetic vector currents. 

\subsection{Lattice calculation}
\begin{table}[tb]
  \begin{center}
    \begin{tabular}{|c | c c c c c c|}
      \hline
      & & & & &\vspace{-0.40cm} \\
      Ensemble & $\beta$ & $a[{\rm fm}]$ & $\left(\frac{L}{a}\right)^3 \times  \frac{T}{a}$ & $m_{PS}$[MeV] &$ L$[fm] & $N_\mathrm{conf}$\\
      & & & & & &\vspace{-0.40cm} \\
      \hline \hline
      & & & & & &\vspace{-0.40cm} \\
      D15.48   & $2.10$ & $0.061$ & $48^3 \times 96$ & 227 & 2.9  & $265/155/156$ \\
      D30.48   & $2.10$ & $0.061$ & $48^3 \times 96$ & 318 & 2.9  & $203/148/148$ \\
      D45.32sc & $2.10$ & $0.061$ & $32^3 \times 64$ & 387 & 1.9  & $397/346/346$ \\
%
      \hline
      & & & & & & \vspace{-0.40cm} \\
      B25.32t& $1.95$ & $0.078$ & $32^3 \times 64$ & 274 &  2.5 & $273/179/180$ \\
      B35.32 & $1.95$ & $0.078$ & $32^3 \times 64$ & 319 &  2.5 & $201/194/194$ \\
      B35.48 & $1.95$ & $0.078$ & $48^3 \times 96$ & 314 &  3.7 & $235/103/104$ \\
      B55.32 & $1.95$ & $0.078$ & $32^3 \times 64$ & 393 &  2.5 & $225/125/125$ \\
      B75.32 & $1.95$ & $0.078$ & $32^3 \times 64$ & 456 &  2.5 & $158/100/100$ \\
      B85.24 & $1.95$ & $0.078$ & $24^3 \times 48$ & 491 &  1.9 & $192/142/136$ \\
      & & & & & &\vspace{-0.40cm} \\
      \hline
      & & & & & &\vspace{-0.40cm} \\
      A30.32 & $1.90$ & $0.086$ & $32^3 \times 64$ & 283 & 2.8 & $267/158/158$ \\
      A40.32 & $1.90$ & $0.086$ & $32^3 \times 64$ & 323 & 2.8 & $248/174/174$ \\
      A50.32 & $1.90$ & $0.086$ & $32^3 \times 64$ & 361 & 2.8 & $216/147/157$ \\
      \hline
    \end{tabular}
    \caption{Parameters of the $N_f = 2+1+1$ flavour gauge field configurations that
      have been analysed in this work. $\beta$ denotes the gauge coupling, $a$ the lattice spacing,
      $\left(\frac{L}{a}\right)^3 \times  \frac{T}{a}$ the space-time volume and $m_{PS}$ is the value of the light pseudoscalar
      meson mass.  The values for $m_{PS}$ have been determined
      in~\protect\cite{Baron:2010bv}. $L$ is
      the spatial extent of the lattices. The lattice spacings are
      taken from~\protect\cite{Baron:2011sf}.
      The last column gives the number of statistically independent gauge configurations used to estimate the
      light/strange/charm contribution to the polarisation function.}
    \label{tab:ensemble_table}
  \end{center}
\end{table}

The strategy for computing and analysing the hadronic vacuum polarisation function is the same as in~\cite{Burger:2013jya,
Burger:2015oya}. In particular, we employ the same set of ensembles~\cite{Baron:2010bv, Baron:2010th}, which is presented in
table~\ref{tab:ensemble_table}. Additionally, we have checked our chiral extrapolations of the light quark contribution by comparing the results with
those obtained on a $N_f=2$ ensemble featuring the physical pion mass~\cite{Abdel-Rehim:2013yaa, Abdel-Rehim:2014nka, Abdel-Rehim:2015pwa}.
The parameters of this ensemble are given in table~\ref{tab:physpoint}.
\begin{table}
 \centering
\begin{tabular}{|c c c c c c c|}
\hline
 & & & & & &\vspace{-0.40cm} \\
 $\beta$ & $c_{\rm SW}$ & $a[{\rm fm}]$ & $\left(\frac{L}{a}\right)^3 \times  \frac{T}{a}$ & $m_{PS}$[MeV] &$ L$[fm] & $N_{\rm conf}$ \\
 & & & & &\vspace{-0.40cm} \\
\hline \hline
& & & & & &\vspace{-0.40cm} \\
 $2.10$ & $ 1.57551$ & $0.094$ & $48^3 \times 96$ & 128 & 4.6 & 804   \\
\hline 
\end{tabular}
\caption{Parameters of ensemble featuring $N_f=2$ twisted mass fermions at the physical point.}
\label{tab:physpoint}
\end{table}
As in~\cite{Burger:2013jya,
Burger:2015oya}, we use
the conserved point-split vector current at
source and sink and  we restrict our considerations to the
quark-connected contributions. In this case, the total vacuum polarisation function 
\begin{equation}
  \Pi^{\rm tot}(Q^2) = \frac{5}{9}  \Pi^{\rm ud}(Q^2) + \frac{1}{9}  \Pi^{\rm s}(Q^2) + \frac{4}{9}  \Pi^{\rm c}(Q^2)
  \label{eq:pi_tot_flavor_sum}
\end{equation}
is obtained by summing the single-flavour contributions which we define without the charge factors.

For each ensemble and each flavour $\rm f$, we first fit the temporal vector current correlator to determine the vector meson
masses, $m_i$, and
their couplings, $g_i$. Then we fit the hadronic vacuum polarisation function obtained from the current correlator as detailed in
\cite{Burger:2013jya}
to the following functional form
\begin{equation}
 \Pi^{\rm f}(Q^2) = 
   (1- \Theta(Q^2-Q^2_{\rm match}))\Pi^{\rm f}_{\mathrm{low}}(Q^2) + \Theta(Q^2-Q^2_{\rm match}) \Pi^{\rm f}_{\mathrm{high}}(Q^2) \;,
\label{eq:pilowandhigh}
\end{equation}
where $\Theta(x)$ is the Heaviside step function. The low-momentum fit function for $0 \le Q^2 \le Q^2_{\rm match}$ is given by
\begin{equation}
  \Pi^{\rm f}_{\mathrm{low}}(Q^2) = \sum_{i=1}^M \frac{g^2_i m^2_i}{m^2_i + Q^2} + \sum_{j=0}^{N-1} a_j (Q^2)^{j} \; ,
\label{eq:pilow}
\end{equation}
and the high-momentum piece for $Q^2_{\rm match}\le Q^2 \le Q_{\rm max}^2$ reads
\begin{equation}
  \Pi^{\rm f}_{\mathrm{high}}(Q^2) = \log(Q^2) \sum_{k=0}^{B-1} b_k (Q^2)^{k}  + \sum_{l=0}^{C-1} c_l (Q^2)^{l} \; . 
\label{eq:pihigh}
\end{equation}
The number of terms and thus the fit function is characterised by M, N, B, and C.
  The ansatz in Eq.~(\ref{eq:pilow}) consists of three parts: a series of poles at energies $m_i$ and with residual $g_i^2 m_i^2$, $i=1,\ldots,M$,
  an additive constant $a_0$ and further polynomial terms $a_i\,( Q^2 )^i$ for $i \ge 1$. The poles characterised by $(m_i,g_i)$ are identified with
the
  exponential contributions to the time-dependent vector-current 2-point correlation function at zero spatial momentum. This is the reason why the
parameters $(m_i,g_i)$ are obtained from a fit of the latter
  and inserted into the fit of the vacuum polarisation function under preservation of all error correlations. The ansatz in Eq.~(\ref{eq:pilow})
  is valid for any four-momentum $Q^2$, in particular for $Q = \left( Q_0,\vec{Q}=0 \right)$ with $Q^2 = Q^2_0$.
  Given such a momentum configuration, the above identification follows from  the Fourier transform of the polarisation tensor.
  With the limited statistical precision of the 2-point vector correlator, the number of exponentials we can resolve in practice is limited to 2.
  Contributions from states with even larger energies are effectively accounted for by the polynomial terms. 

  The ansatz in Eq.~(\ref{eq:pihigh}) is chosen to provide an adequate parametrisation of the polarisation function. 
  This is the only requirement in the high-momentum region $Q^2_\mathrm{match} < Q^2 \le Q^2_\mathrm{max}$. While in the low-momentum region
  the extrapolation beyond the lowest non-zero lattice momentum to zero momentum is of physical significance as it predicts the curvature of
  the polarisation function in this interval, in the high-momentum region we only need the ansatz in Eq.~(\ref{eq:pihigh}) to interpolate the
available lattice data.

  In the following, we use $Q^2_{\rm max}= 100\,{\rm GeV}^2$ and choose $Q^2_{\rm match}=2\,{\rm GeV}^2$. 
  Varying $Q^2_{\rm match}$ by $1\,{\rm GeV}^2$ to the left and to the right gives compatible results.
  We perform extrapolations of the subtracted polarisation function in the light quark mass and the lattice spacing only for momenta in
  the interval $0 < Q^2 \le 10\,{\rm GeV}^2$ since for larger $Q^2$ perturbative calculations are expected to yield more precise determinations. 

\begin{figure}[htb]
 \centering 
\hfill
\includegraphics[width=0.47\textwidth]{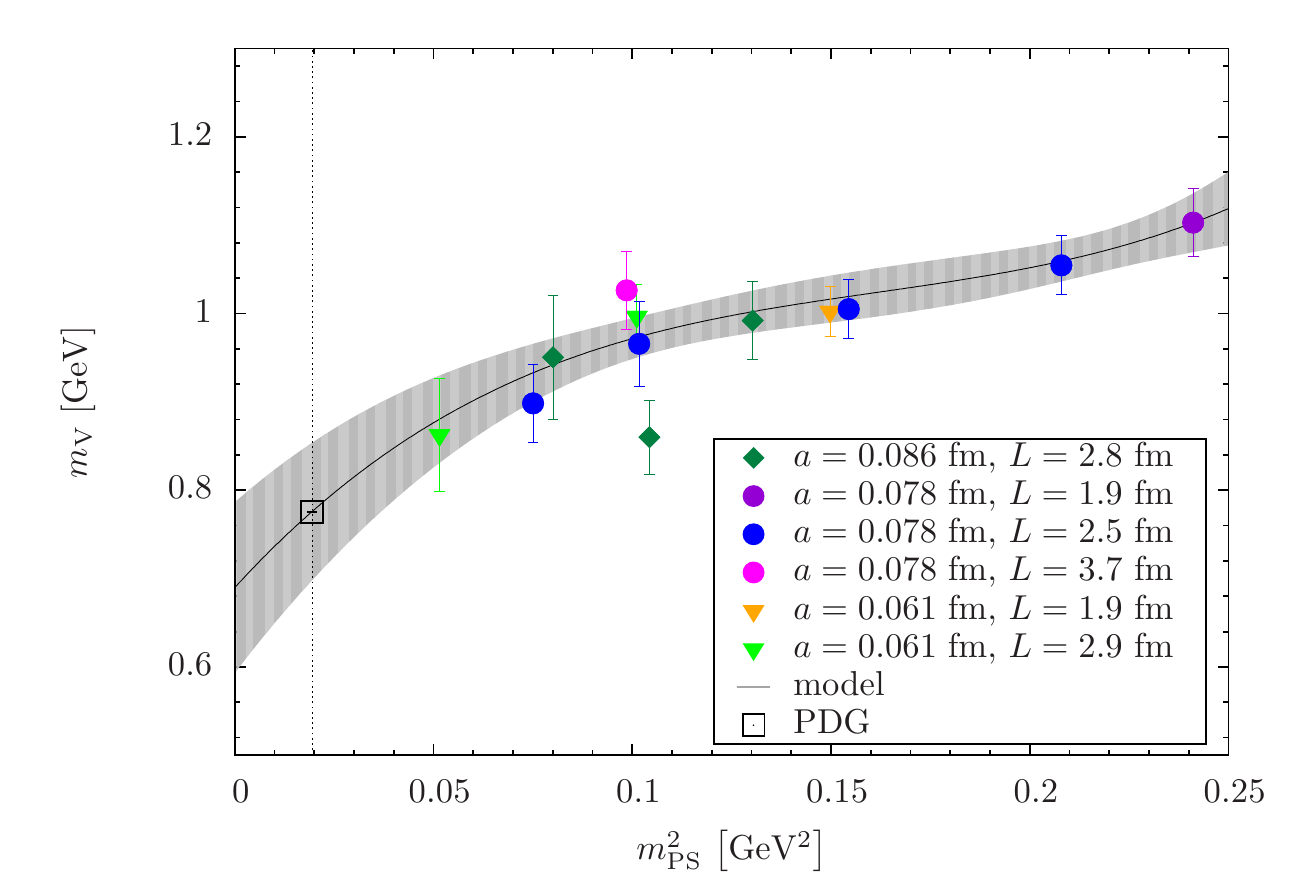}
\hfill
\includegraphics[width=0.47\textwidth]{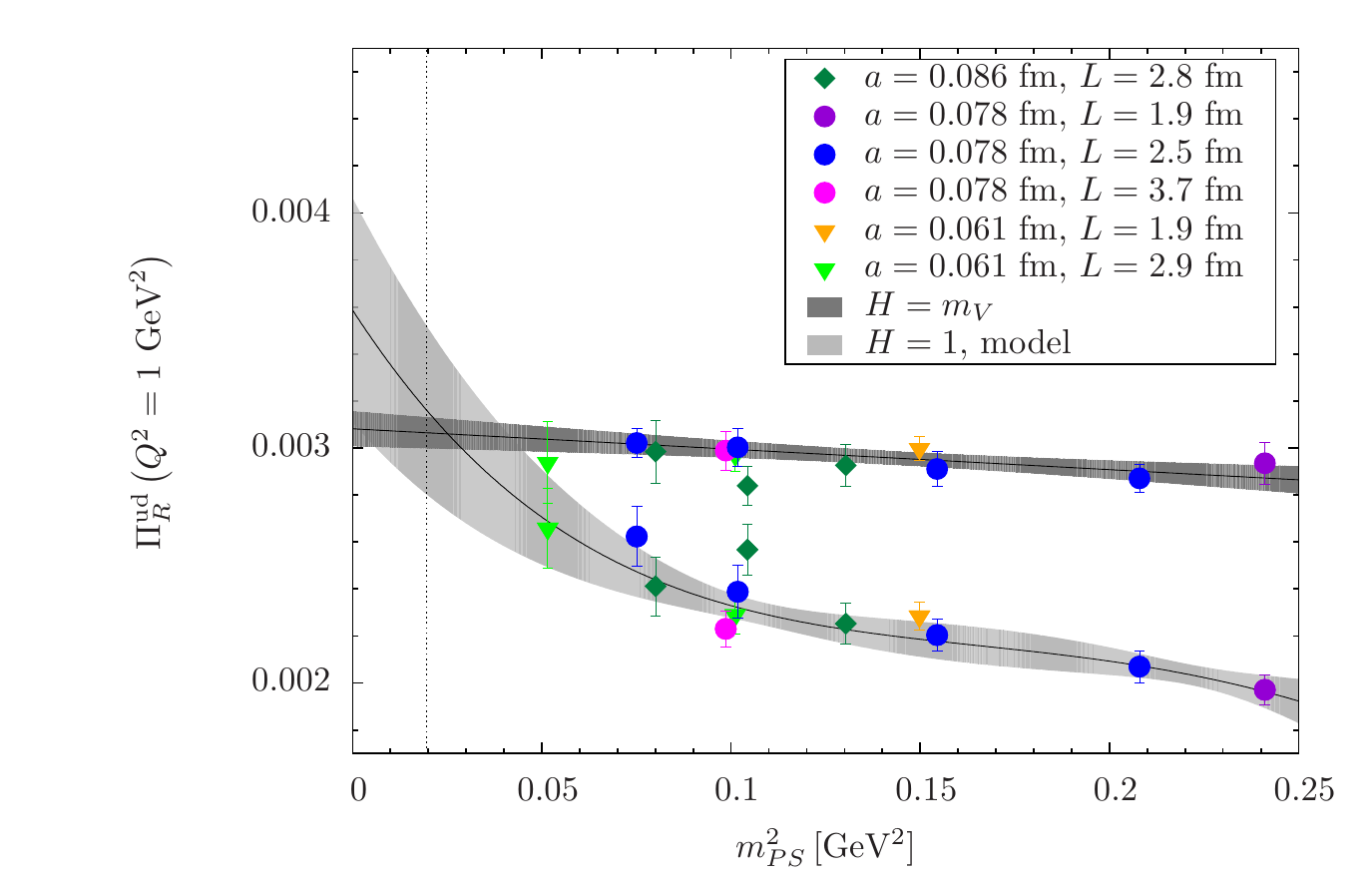}
\hfill
\caption{Left: Dependence of $\rho$-meson mass, $m_V$, on the squared pion mass, $m_{\rm PS}^2$, determined on the $N_f=2+1+1$ twisted mass ensembles shown in table \ref{tab:ensemble_table}. Right: Chiral extrapolation of the renormalised light-quark vacuum polarisation function at a generic value of $Q^2 = 1\,{\rm GeV}^2$ obtained with the redefinition given in Eq.~(\ref{eq:pi_redef}) (upper set of data points) and with the standard definition Eq.~(\ref{eq:Pi_R}) (lower set of data points). For the latter the same model extrapolation as for the $\rho$-meson mass has been used.}
\label{fig:mv}
\end{figure}

Since our four-flavour ensembles feature unphysically large pion masses, an extrapolation to the physical point has to be
performed. The pion mass dependence of the single-flavour contributions can be assessed by looking at the leading vector meson contribution obtained
in chiral perturbation theory~\cite{Ecker:1988te, Aubin:2006xv}
\begin{equation}
\Pi^{\rm f}(Q^2) = g_V^2 \frac{m_V^2}{Q^2+m_V^2} \; .
\label{eq:Pi_chipt}
\end{equation}
 The spectral properties of the heavy vector mesons hardly depend on the pion mass and also the coupling constant $g_V$ of the $\rho$-meson has been found to be
well-described by a
linear fit
in the squared pion mass, $m_{\rm PS}^2$, cf.~\cite{Renner:2012fa}. However, the $\rho$-meson mass, $m_V$, strongly depends on the value of the light quark
masses, taken to be degenerate in our calculation, and thus the squared pion mass~\cite{Burger:2013jya}. This is illustrated in
Fig.~\ref{fig:mv} left by a model extrapolation which is constrained by requiring the $\rho$-meson mass to attain its experimental value~\cite{Agashe:2014kda} at the physical point, similarly to the one used in the two-flavour case in Ref.~\cite{Feng:2011zk}.

\Eq.~(\ref{eq:Pi_chipt}) implies that a similar non-linear behaviour can be expected for the light-quark hadronic vacuum polarisation function as we indeed observe for the lower set of data points in Fig.~\ref{fig:mv} right.
From Eq.~(\ref{eq:Pi_chipt}) we also see that we can eliminate this non-linear dependence on the squared pion mass to a large extent by employing the
lattice redefinition presented in Ref.~\cite{Renner:2012fa} for the light-quark contribution to the vacuum polarisation function,
\begin{equation}
\label{eq:pi_redef}
  \bar{\Pi}_R^{\rm ud}\left( Q^2 \right) = \Pi^{\rm ud}_R\left( Q^2 \cdot \frac{H^2}{H_\mathrm{phys}^2} \right)\,,
\end{equation}
if we use $H=m_V$, i.e. the $\rho$-meson mass at unphysically large up and down quark masses. The beneficial effect this has on the data points is depicted as the upper set of data points in Fig.~\ref{fig:mv} right. Hence, in the following we use the above redefinition in the light-quark sector.
For the contributions of the heavy quark flavours, we use the standard definition of the vacuum polarisation function $\Pi^{\rm f}_R(Q^2)$,
${\rm f} = {\rm s,c}$.
Our redefinition of the total $\Delta \alpha_{\mathrm{QED}}^{\mathrm{hvp}}$ then follows from the sum over all quark flavours as in
Eq.~(\ref{eq:pi_tot_flavor_sum})
\begin{equation}
  \Delta \overline{\alpha}_{\mathrm{QED}}^{\mathrm{hvp}}(Q^2) = -4 \pi \alpha_0 \,\left( 
    \frac{5}{9}\,\Pi^{\rm ud}_{\mathrm{R}}\left(Q^2 \cdot \frac{ H^2}{ H_{\mathrm{phys}}^2}\right) +
    \frac{1}{9}\,\Pi^{\rm s}_{\mathrm{R}}\left(Q^2 \right) +
    \frac{4}{9}\,\Pi^{\rm c}_{\mathrm{R}}\left(Q^2 \right) 
  \right)\,.
  \label{eq:delta_alpha_redefined}
\end{equation}
The lattice data obtained with the definition $\Delta \overline{\alpha}_{\mathrm{QED}}^{\mathrm{hvp}}$ in Eq.~(\ref{eq:delta_alpha_redefined})
can be sufficiently well described already by a linear dependence on the squared pion mass. Since we use only this definition
throughout this work
and no confusion is possible, we henceforth omit the bar in $\Delta \overline{\alpha}_{\mathrm{QED}}^{\mathrm{hvp}}$. 

\subsection{Results}

\begin{figure}[htb]
 \centering
\hfill
\includegraphics[width=0.5\textwidth]{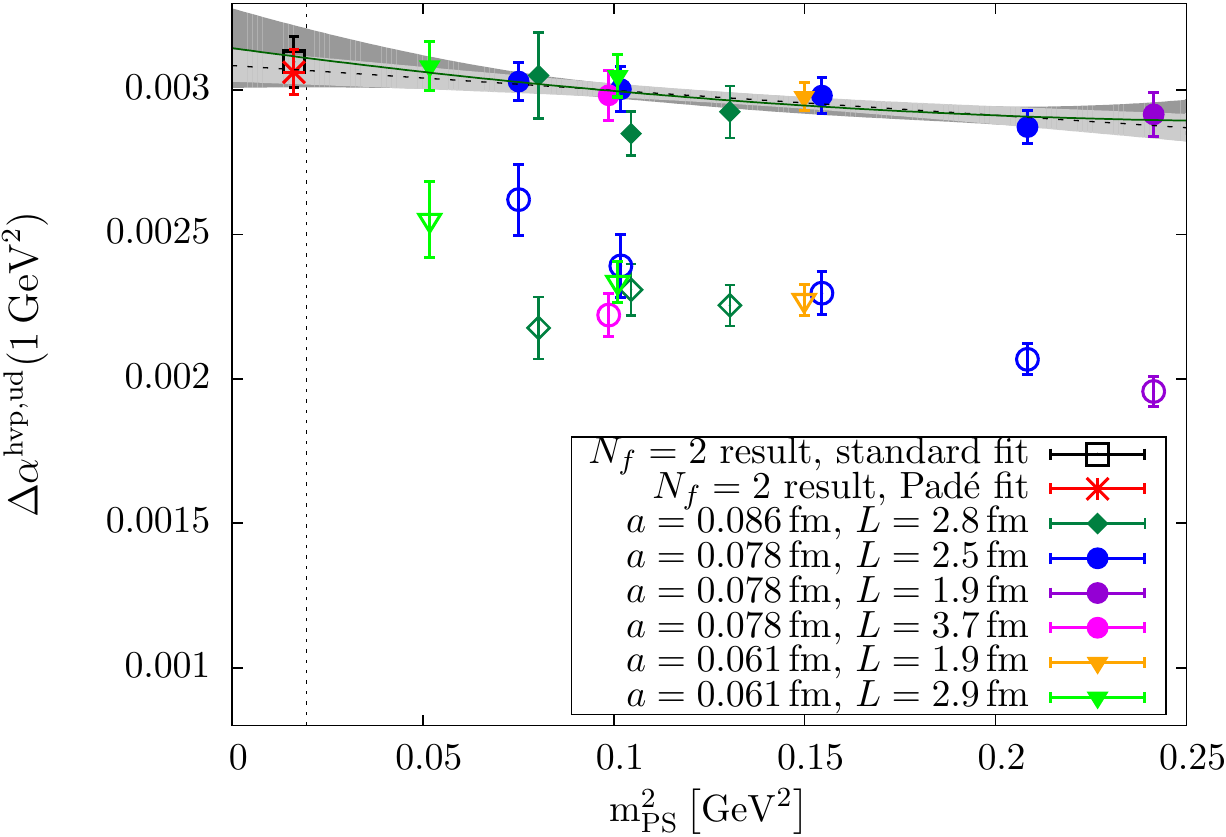}
\hfill
\includegraphics[width=0.4\textwidth]{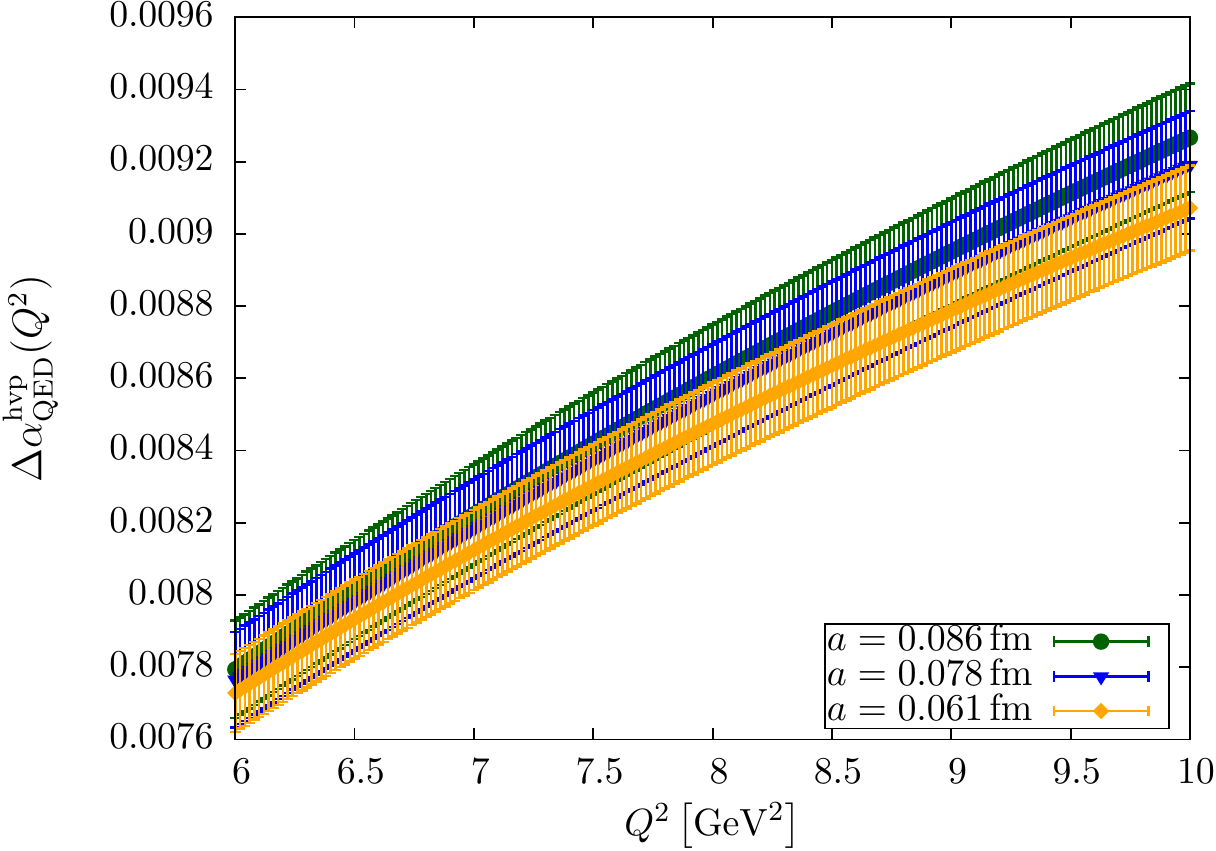}
\hfill
\caption{Left: Light-quark contribution to $\Delta \alpha_{\rm
QED}^{\rm hvp}$ with filled symbols representing points obtained with
Eq.~(\ref{eq:pi_redef}) using $H=m_V$, open symbols
refer to those obtained with  Eq.~(\ref{eq:def_alpha}), i.~e.~$H=1$ in Eq.~(\ref{eq:pi_redef}). In particular, the two-flavour results at the
physical point have been computed with the standard
definition. The light grey errorband displays the uncertainty of the linear fit represented by the black dotted line, wheras the dark grey errorband
belongs to a quadratic fit represented by the green solid line. Right: $N_f=2+1+1$ contribution to
$\Delta \alpha_{\rm QED}^{\rm hvp}$ 
for the three lattice spacings at a fixed pion mass of $m_{\rm PS} \approx 320\, {\rm MeV}$.}
\label{fig:alpha_light_heavy}
\end{figure}

In order to show that the above redefinition in Eq.~(\ref{eq:pi_redef}) indeed provides the expected benefit for the chiral
extrapolation of the
light quark contribution to the running of the fine structure constant, we show the data for both Eqs.~(\ref{eq:def_alpha}) and
(\ref{eq:pi_redef}) with $H=m_V$
in Fig.~\ref{fig:alpha_light_heavy} left for a single momentum value $Q^2=1\,{\rm GeV}^2$. The upper set of data points obtained with the redefinition
Eq.~(\ref{eq:pi_redef}) evidently is much easier to extrapolate to the physical value of the pion mass than the lower points procured from the
standard definition Eq.~(\ref{eq:def_alpha}).

Since we do neither observe lattice spacing artefacts nor finite size effects in these
data at $Q^2=1\,{\rm GeV}^2$, we can actually compare our results computed on the four-flavour ensembles linearly extrapolated in the squared pion
mass,
$m_{\rm PS}^2$, with those obtained from the $N_f=2$ ensemble featuring the physical pion mass. 
Additionally to the standard analysis, we have
performed a correlated $[1,1]$ Pad\'e fit \cite{Aubin:2012me}, possessing the same number of parameters, up to $Q^2_{\rm max} =1.5\, {\rm GeV}^2$ such
that $Q^2=1\,{\rm GeV}^2$ is safely
covered. As expected, the values for the pole parameters determined from the temporal
correlator in our standard approach and from the Pad\'e fit are compatible
\begin{equation}
 a^2\,m_V^2 = 0.153(35) \qquad b_n = 0.1575(81)
\end{equation}
and also the results of both analyses of the leading hadronic contribution to the running of the fine structure constant at the physical point
completely agree with each other and with the extrapolated result obtained on the four-flavour ensembles indicating that the systematic uncertainty
caused by the chiral extrapolation is small. The results at the physical value of the pion mass, which are depicted in the left panel of
Fig.~\ref{fig:alpha_light_heavy}, are
summarised in table~\ref{tab:alpha_light}.
\begin{table}[htb]
\begin{center}
\begin{tabular}{|c|cc|}
 \hline
 $N_f=2+1+1$ extrapolated & $N_f=2$ standard & $N_f=2$ $[1,1]$ Pad\'e\\
 \hline
 \hline
  0.003068(50) &  0.003097(88) & 0.003062(77)\\
 \hline
\end{tabular}
\end{center}
\caption[Comparison of the chirally extrapolated result for $\Delta \alpha_{\rm QED}^{\rm hvp, ud}(1\,{\rm GeV}^2)$ obtained on the $N_f=2+1+1$
ensembles
with those
obtained on the $N_f=2$ ensemble at the physical point.]{Comparison of the chirally extrapolated result for $\Delta \alpha_{\rm QED}^{\rm hvp,
ud}(1\, {\rm GeV}^2)$ obtained on the $N_f=2+1+1$ ensembles with those obtained on the $N_f=2$ ensemble at the physical point. For the latter, we have
performed our standard analysis but without the redefinition and also tested a [1,1] Pad\'e fit.}
\label{tab:alpha_light}
\end{table}

Including also the heavy quark contributions by using Eq.~(\ref{eq:delta_alpha_redefined}), a
dependence on the lattice spacing is clearly visible, especially in the high-$Q^2$ region shown in the right panel of
Fig.~\ref{fig:alpha_light_heavy}.
This is  accounted for by combining
the chiral extrapolation with taking the
continuum limit and employing
the following fit function to the four-flavour results obtained on individual ensembles

\begin{equation}
 \Delta \alpha_{\mathrm{QED}}^{\mathrm{hvp}}(Q^2)(m_{\rm PS}, a)= A+ B~m_{\rm PS}^2 + C~a^2
\label{eq:extrap}
\end{equation}
with fit parameters $A$, $B$, $C$ for each momentum value $Q^2 \in \{ 0, 0.02, 0.04, \ldots, 10\}\, {\rm GeV}^2$. In~\cite{Burger:2014ada} we have
shown that automatic $\mathcal{O}(a)$ improvement is at work for our definition of the hadronic vacuum polarisation function. Thus, performing the
continuum extrapolation without a term linear in the lattice spacing $a$ in
Eq.~(\ref{eq:extrap}) is justified. The ansatz in Eq.~(\ref{eq:extrap}) neglects any dependence on the finite lattice extent which
has been found to be smaller than our current statistical uncertainties and will be discussed when assessing the systematic effects of our calculations.

\begin{figure}[htb]
 \centering
\includegraphics[width=0.65\textwidth]{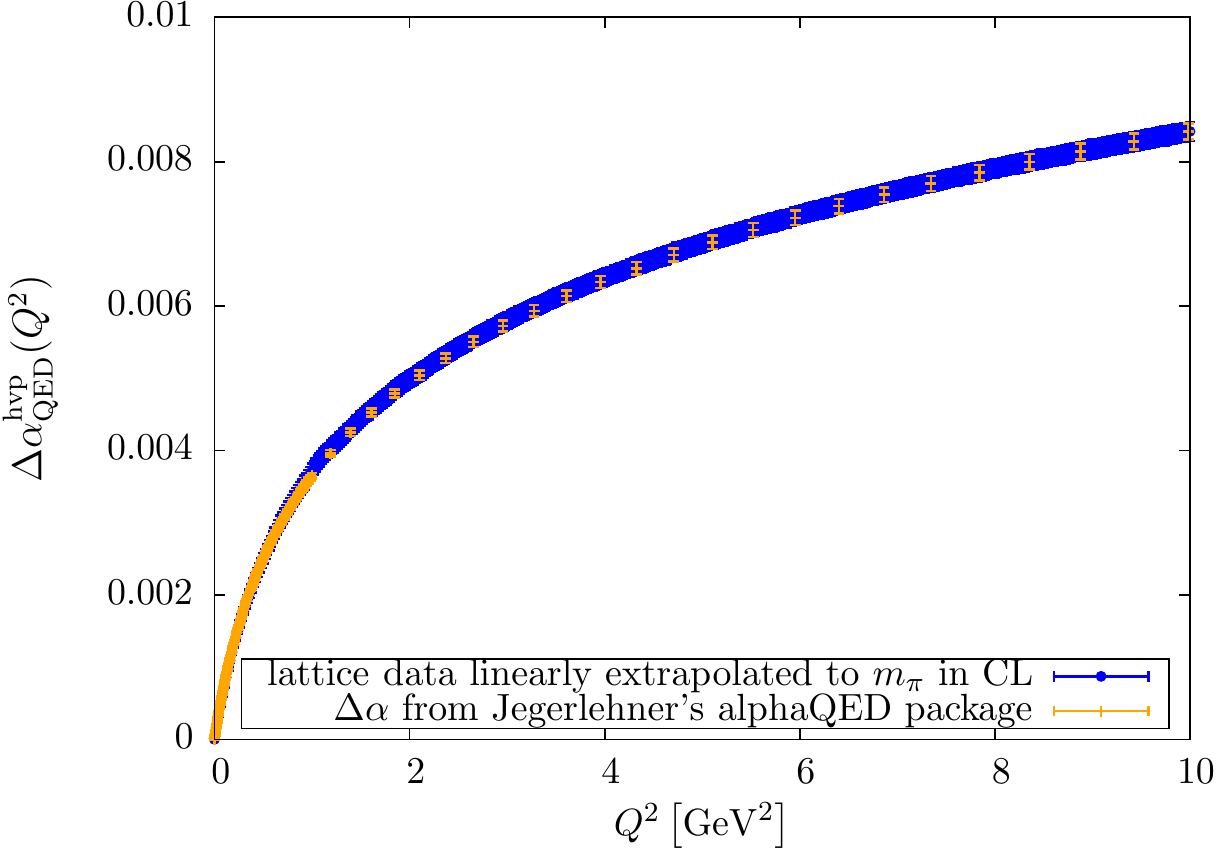}
\caption{$N_f=2+1+1$ contribution to $\Delta \alpha_{\rm QED}^{\rm hvp}$ 
compared to the data collected in~\cite{Jegerlehner:2012:Online} employing
the
dispersion
relation in Eq.~(\ref{eq:disp}). The lattice data are taken at the physical point and in 
the continuum limit (CL).}
\label{fig:alpha_tot}
\end{figure}

The
results are depicted in
Fig.~\ref{fig:alpha_tot} together with the results obtained by a phenomenological analysis~\cite{Jegerlehner:2011mw}. Here, for
both the lattice calculation and the phenomenological analysis only the statistical errors are shown. Over the whole momentum
range, perfect agreement with comparable statistical uncertainties is found. 
We will discuss the systematic uncertainties of our lattice QCD determination below. An updated phenomenological analysis including
all data published till the end of 2014 will soon be
available~\cite{Jegerlehner:2015}. The lattice data also agree with those results featuring even smaller uncertainties.

\subsubsection{Systematic uncertainty from the choice of vector meson fit ranges}
\begin{figure}[htb]
 \centering
\includegraphics[width=0.55\textwidth]{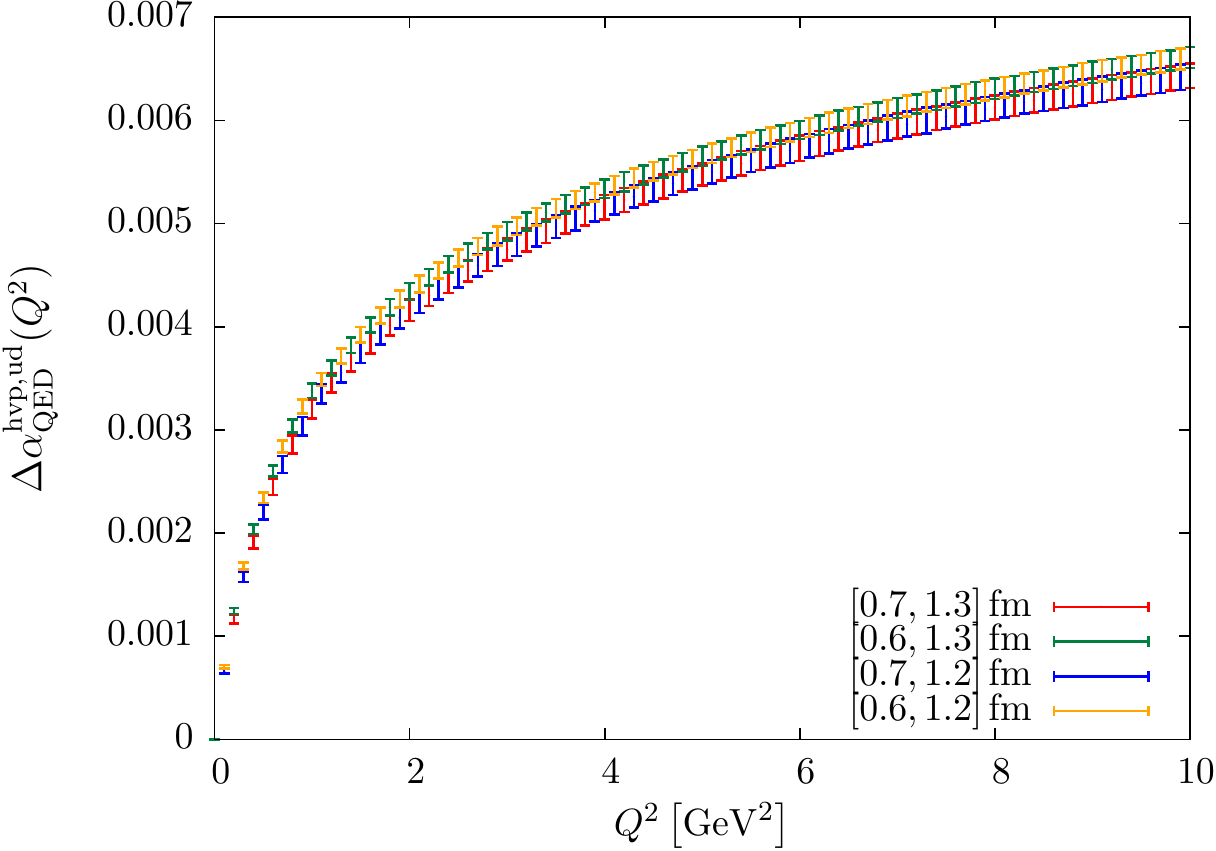}
\caption{Light quark contribution to 
$\Delta \alpha_{\rm QED}^{\rm hvp}$ obtained with different fit ranges for the 
$\rho$-meson mass, $m_V$ and coupling, $g_V$. The
standard fit range is $[0.7\, {\rm fm},1.2\,{\rm fm}]$.}
\label{fig:alpha_light_fitranges}
\end{figure}
As mentioned before, the first step in our analysis is the determination of the masses and the coupling constants of the vector mesons from the
vector two-point functions at zero momentum. The values of the spectral parameters differ when varying the fit range. We have repeated the complete
analysis
for various vector meson fit ranges for the light, strange and charm quark currents propagating the uncertainty to the final results.

In the light quark sector depicted in Fig.~\ref{fig:alpha_light_fitranges}, we observe systematic uncertainties depending on whether we start fitting
the vector meson correlator at $0.6\,{\rm fm}$ or at $0.7\,{\rm fm}$ whereas changing the upper border of the fit interval by $0.1\, {\rm fm}$ does
not lead to observable effects. The dependence on the lower starting point of the 
fit can be attributed to excited state contamination of the $\rho$-meson correlator. When stating the final results
for selected
momentum values below, we take for these systematic uncertainties half the difference between the central
values that are furthest apart from each other.

\begin{figure}[htb]
 \centering
 \hfill
\includegraphics[width=0.45\textwidth]{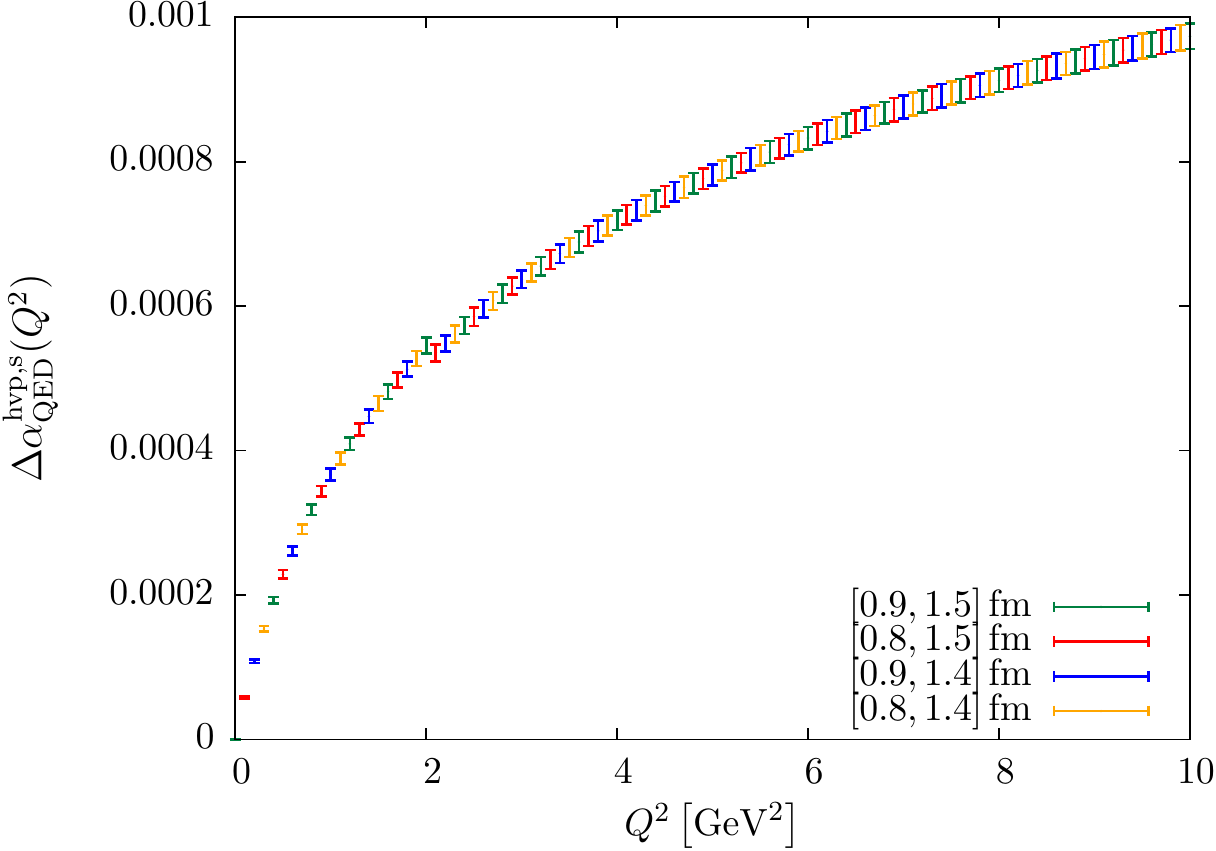}
\hfill
\includegraphics[width=0.45\textwidth]{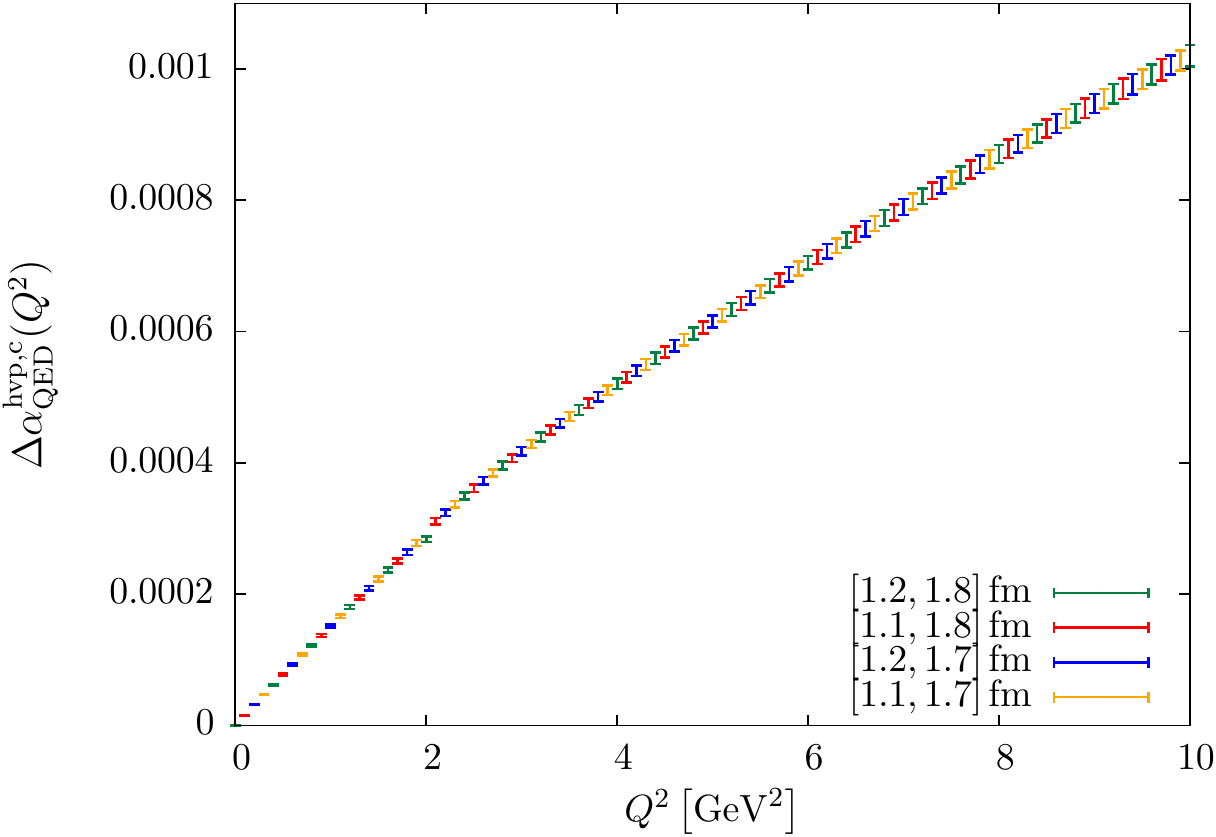}
\hfill
\caption{Dependence of the single-flavour contributions to $\Delta \alpha_{\rm QED}$ on the fit range of the $\overline{s}s$-correlator (left panel)
and
of the $J/\Psi$-correlator (right panel). 
The standard $\overline{s}s$-correlator fit range is $[0.9\, {\rm fm},1.4\,{\rm fm}]$, whereas the one for the charm quark correlator is $[1.2\, {\rm
fm},1.7\,{\rm fm}]$. The minor discontinuities at $Q^2= 2\,{\rm GeV}^2$ arise from connecting the low-momentum Eq.~(\ref{eq:pilow}) and high-momentum
Eq.~(\ref{eq:pihigh}) fit functions at this point by a simple step function as shown in Eq.~(\ref{eq:pilowandhigh}). Due to the subdominance of the
heavy flavour contributions, those discontinuities do not influence the final result.}
\label{fig:alpha_heavy_fitranges}
\end{figure}

For the heavy flavours changing the fit interval by $0.1\, {\rm fm}$ to the left and to the right of both the lower and the upper time slice of the
fit ranges does not lead to observable differences. This is shown in Fig.~\ref{fig:alpha_heavy_fitranges}.

\subsubsection{Systematic uncertainty from the choice of vacuum polarisation fit function}
\begin{figure}[htb]
 \centering
\includegraphics[width=0.55\textwidth]{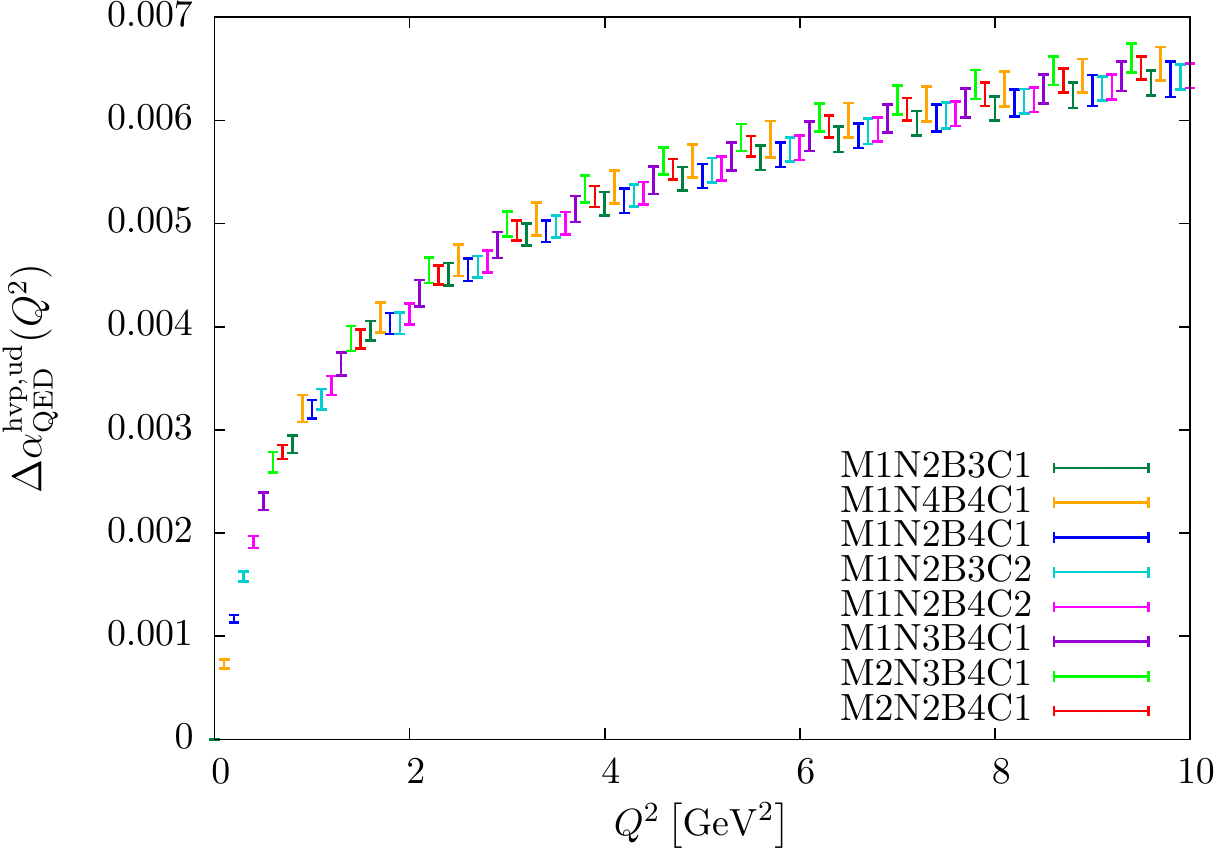}
\caption{Light quark contribution to $\Delta \alpha_{\rm QED}^{\rm hvp}$ obtained from different fit functions. The
standard fit is M1N2B4C1.}
\label{fig:alpha_light_mnbc}
\end{figure}

Performing the whole analysis with different numbers of terms in our vacuum polarisation fit functions also leads to observable differences in the
light
quark contribution as shown in Fig.~\ref{fig:alpha_light_mnbc}. These are larger than the effects from the fit ranges of the vector meson fits
discussed in the preceding subsection and thus present the dominant systematic uncertainty in our calculation. It might be possible to improve the situation by  
 e.g. the method of analytic continuation~\cite{Feng:2013xsa, Francis:2013fzp} or by taking
momentum derivatives of the vacuum polarisation function~\cite{deDivitiis:2012vs}.

\begin{figure}[htb]
 \centering
 \hfill
\includegraphics[width=0.45\textwidth]{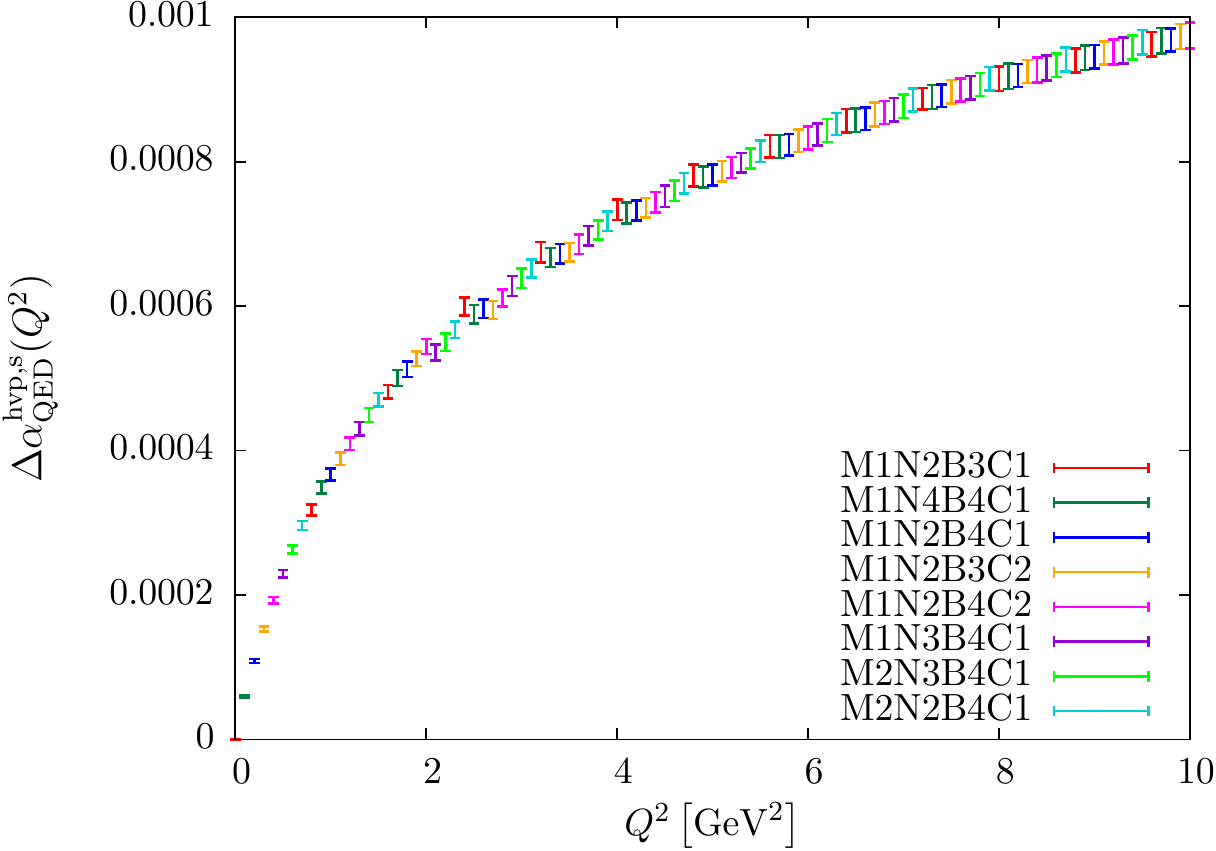}
\hfill
\includegraphics[width=0.45\textwidth]{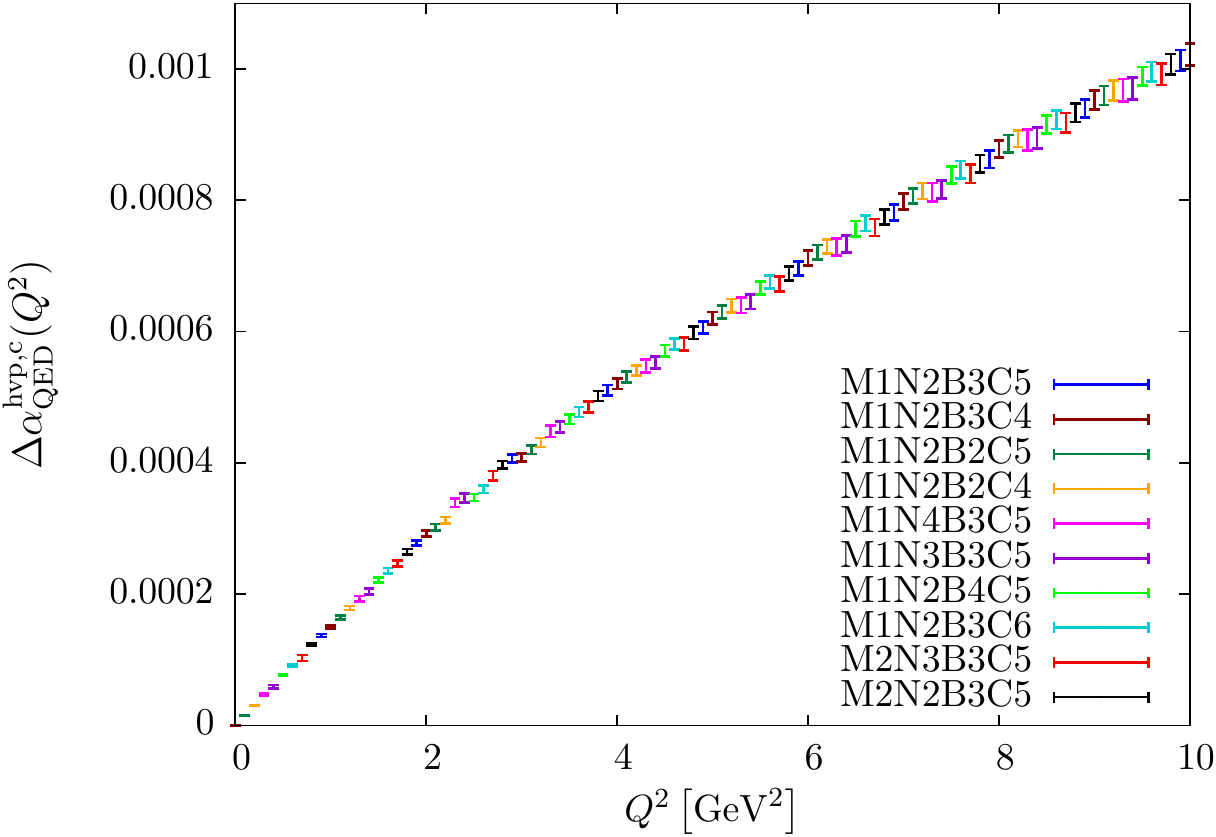}
\hfill
\caption{Dependence of the single-flavour contributions to $\Delta \alpha_{\rm QED}$ on the choice of fit function for the strange (left panel) and
for the charm (right panel) quark pieces. 
For the strange quark the standard fit is M1N2B4C1, whereas the one for the charm quark correlator is M1N2B3C5.}
\label{fig:alpha_heavy_mnbc}
\end{figure}

The situation for the heavy quarks is shown in Fig.~\ref{fig:alpha_heavy_mnbc}. Here, almost no systematic deviations are visible. Furthermore the
contributions from the heavy quarks are about an order of magnitude smaller than the light-quark one. Hence, we do not take systematic effects from the
variation of the second-generation quark fit functions into account in our final error estimate.

\subsubsection{Finite size effects}
\begin{figure}[htb]
 \centering
 \hfill
\includegraphics[width=0.45\textwidth]{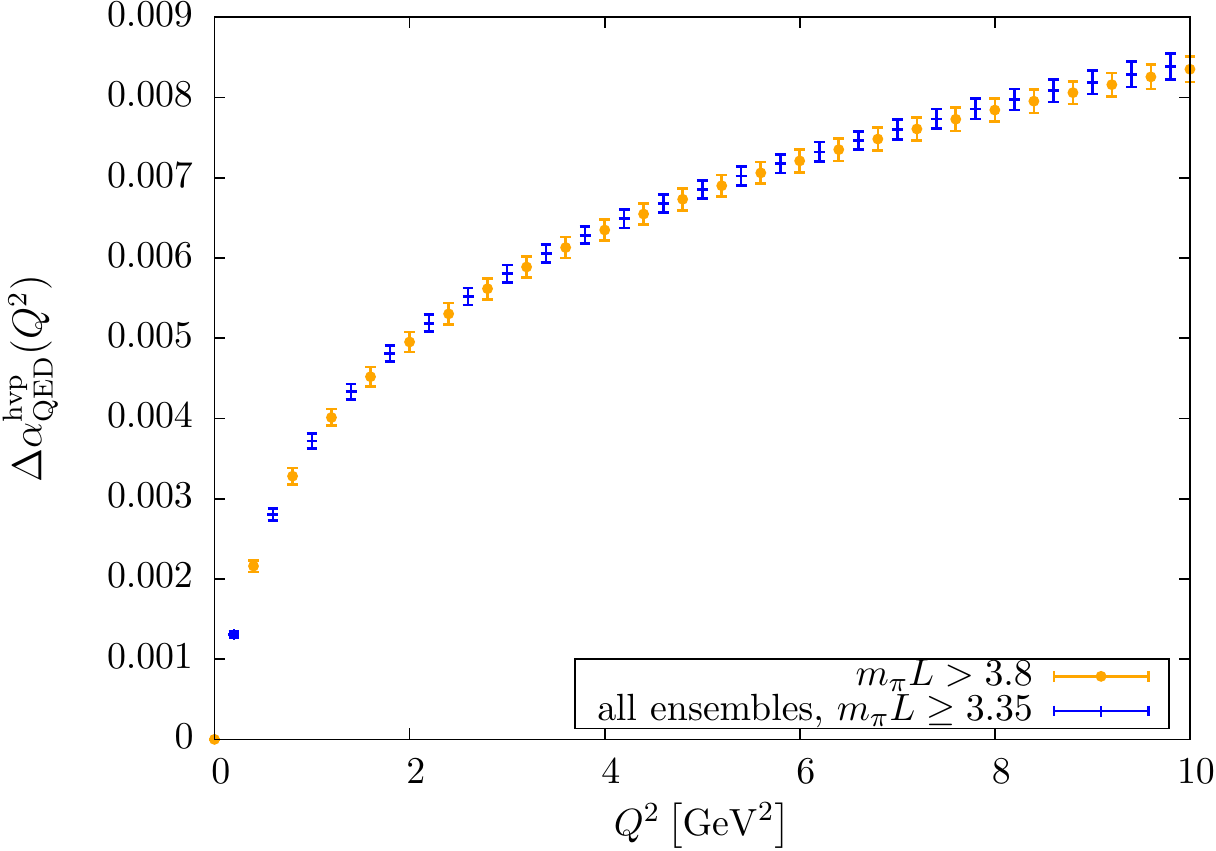}
 \hfill
\includegraphics[width=0.45\textwidth]{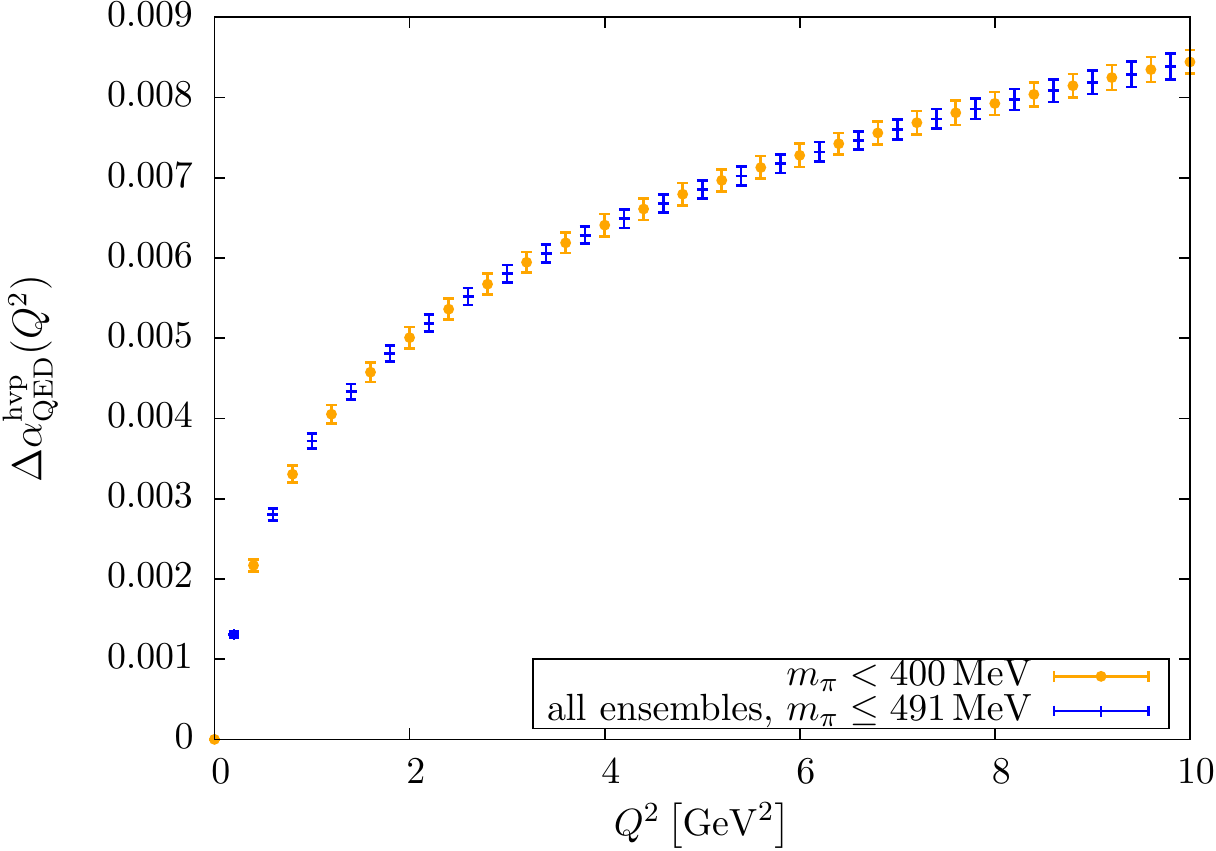}
\caption{Four-flavour contribution to $\Delta \alpha_{\rm QED}^{\rm hvp}$ obtained with (left panel) $m_{\rm PS} L \ge 3.35$ [standard] and $m_{\rm
PS} L > 3.8$ and (right panel) $m_{\rm PS} \le 491\,{\rm MeV}$ [standard] and $m_{\rm PS} L
 < 400\,{\rm MeV}$.}
\label{fig:alpha_FSE_mpi}
\end{figure}

In lattice QCD, typically $m_\mathrm{PS}~L \gtrsim 4$ is required to minimise systematic
effects due to the finite lattice volumes, where $L$ denotes the spatial extent of the lattice.
The $N_f=2+1+1$ ensembles analysed in this work feature 
$3.35 < m_\mathrm{PS}~L < 5.93$. 
Restricting our data to the condition $m_{\rm PS} L > 3.8$ yields the picture shown in the left panel of Fig.~\ref{fig:alpha_FSE_mpi}. Hence, we do
not
associate a systematic uncertainty to the usage of ensembles possessing smaller $m_{\rm PS} L$ values. 

\subsubsection{Systematic uncertainty from including heavy pion masses}
In order to extrapolate to the physical point, $m_\pi \approx 140\,{\rm MeV}$, often not too high pion masses should be included in the fit. The ensembles
entering the standard analysis comprise pion masses up to $m_{\rm PS} \approx 491\,{\rm
MeV}$. Using only the ensembles with $m_{\rm PS} < 400\,{\rm
MeV}$ yields fully compatible results for $\Delta \alpha_{\rm QED}^{\rm
hvp}$ as can be seen in the right panel of Fig.~\ref{fig:alpha_FSE_mpi}. Therefore, we do not account for a systematic uncertainty related to the
usage of pion masses above $400\,{\rm MeV}$.

\subsubsection{Final results for selected momentum values}
Table~\ref{q2qed} 
contains our final results compared to those of a phenomenological analysis~\cite{Jegerlehner:2011mw} utilising the once-subtracted
dispersion relation Eq.~(\ref{eq:disp}). The first error denotes the statistical and the second error the systematic uncertainty of our results.
The latter constitutes the dominant source of uncertainty of which the biggest part originates from the choice of the vacuum polarisation fit
function. This might change when lowering the statistical uncertainty, because then the vacuum polarisation fit gets more constrained. Alternatively, avoiding to fit the vacuum polarisation might be considered.

\begin{center}
\begin{table}[tb]
\centering
 \begin{tabular}{c | c c }
 \hline 
  & \vspace{-0.40cm} \\
 $Q^2$ [$\rm GeV^2$]& this work & dispersive analysis~\cite{Jegerlehner:2011mw} \\
 \hline \hline
 0.02 & $0.163\,(05)\,(09)\cdot 10^{-3}$  & $0.174\,(02)\cdot 10^{-3}$\\
 1.00 & $3.721\,(96)\,(145)\cdot 10^{-3}$  & $3.651\,(40)\cdot 10^{-3}$\\
 2.00 & $4.993\,(102)\,(144)\cdot 10^{-3}$ & $4.916\,(61)\cdot 10^{-3}$\\
 3.00 & $5.800\,(111)\,(151)\cdot 10^{-3}$ & $5.725\,(74)\cdot 10^{-3}$\\
 4.00 & $6.396\,(108)\,(156)\cdot 10^{-3}$ & $6.333\,(84)\cdot 10^{-3}$\\
 6.00 & $7.264\,(114)\,(159)\cdot 10^{-3}$ & $7.223\,(98)\cdot 10^{-3}$\\
 8.00 & $7.906\,(124)\,(151)\cdot 10^{-3}$ & $7.850\,(107)\cdot 10^{-3}$\\
 10.0 & $8.419\,(130)\,(159)\cdot 10^{-3}$ & $8.420\,(114)\cdot 10^{-3}$\\
 \end{tabular}
\label{q2qed}
\caption{We tabulate $\Delta \alpha_{\rm QED}^{\rm hvp}(Q^2)$ for selected 
values of $Q^2$. The first 
error of the lattice results is statistical, the second systematic. 
The phenomenological values of $\Delta \alpha_{\rm QED}^{\rm hvp}(Q^2)$ 
have been obtained from the dispersive analysis of Ref.~\cite{Jegerlehner:2011mw}.}
\end{table}
 \end{center}

\section{The weak mixing angle $\sin^2\theta_W$} 
\label{sec:theta}
The weak mixing or Weinberg angle, $\theta_W$, is one of the fundamental parameters of the electroweak standard model defined by
  \begin{equation}
   \sin^2 \theta_W= \frac{g'^2}{g'^2+g^2} = \frac{e^2}{g^2}= \frac{\alpha_{\rm QED}}{\alpha_2}
  \end{equation}
where  $g$ is the $SU(2)_L$ coupling constant and $g'$ the $U(1)_Y$ coupling constant. The second equality is the electroweak unification
condition $e^2 =g^2 \sin^2\theta_W$ for the positron charge $e$. Thus, the running of the weak mixing angle can be
obtained from the running of the fine structure constant and the $SU(2)_L$ coupling
$\alpha_2$.
In the leading logarithmic approximation this is given by~\cite{Jegerlehner:1986vs}
\begin{equation}
 \sin^2\theta_W(Q^2) = \sin^2(\theta^0)\frac{1-\Delta\alpha_2 (Q^2)}{1-\Delta \alpha_{\rm QED}(Q^2)} = \sin^2(\theta_0) (1+ \Delta(Q^2))
\end{equation}
where $\sin^2(\theta_0)=\frac{\alpha^0}{\alpha_2^0}$, and $\Delta(Q^2)= \Delta \alpha_{\rm QED}(Q^2) -\Delta \alpha_2(Q^2)$ is an abbreviation for
$\Delta\sin^2\theta_W(Q^2)$.
The value of 
$\sin^2(\theta_0)$ has essentially been measured by the Boulder group studying atomic parity violation in Cesium~\cite{Wood:1997zq}, the latest value
is $\sin^2(\theta_0)=0.2356(20)$~\cite{Dzuba:2012kx}. The standard model prediction in the $\overline{\rm MS}$ scheme is
$\sin^2(\theta_0)=0.23871(9)$~\cite{Erler:2004in, Kumar:2013yoa} which is
the value employed in the analysis below in order to gain fully theoretical results without experimental input. In the computation of this value the
Higgs boson mass determined by the LHC experiments~\cite{Aad:2012tfa, Chatrchyan:2012ufa} has been used.

A phenomenological value of the leading hadronic contribution to the running of
$\sin^2\theta_W$ between $0$ and the Z-scale has been computed for the first time in~\cite{Marciano:1993jd} relying on results
of~\cite{Jegerlehner:1991dq}. The method
has been described and used with an older dispersive analysis~\cite{Wetzel:1981vt} before~\cite{Marciano:1983ss}. In~\cite{Erler:2004in} the error
has been reduced with respect to the original rather conservative estimate of the uncertainty by about an order of magnitude.

The leading hadronic contribution to the running of the $SU(2)_L$ coupling constant originates from Z-$\gamma$ mixing
\begin{center}
  \begin{tikzpicture}
   \draw[dashed, thick] (0.5,0)--node[below] {Z} (2,0);
   \filldraw[lightgray] (2,0) circle (10 pt);
   \node[circle] at (2,0) {had};
   \draw[decorate, decoration={snake}] (2.3,0)--node[below] {$\gamma$} (3.8,0);
  \end{tikzpicture}
\end{center}
From the expressions for the hadronic currents of up-type ($u$) and down-type ($d$) quarks
\begin{eqnarray}
 J_{\mu}^Z &=& J_{\mu}^3 -\sin^2(\theta_W) J_{\mu}^{\gamma} \\
 J_{\mu}^3 &=& \frac{1}{4} \sum_f \left(\overline{u}_f \gamma_{\mu} (1-\gamma_5) u_f -\overline{d}_f \gamma_{\mu} (1-\gamma_5) d_f\right) \\
 J_{\mu}^{\gamma} & = & \sum_f \left(\frac{2}{3}
\overline{u_f} \gamma_{\mu} u_f -\frac{1}{3}\overline{d_f} 
\gamma_{\mu} d_f\right)
\end{eqnarray}
where 3 refers to the third component of the weak isospin current and $\gamma$ to the
electromagnetic current discussed already above,
we see that to leading order
\begin{equation}
 \Pi^{Z \gamma} \approx \Pi^{3\gamma} = \langle J_{\mu}^3  J_{\mu}^{\gamma} \rangle
\label{eq:pi_zgamma}
\end{equation}
and thus the leading hadronic contribution to the running of $\alpha_2$ is given by~\cite{Jegerlehner:1985gq, Jegerlehner:2011mw}
\begin{equation}
 \Delta \alpha_2^{\rm hvp} (Q^2)= - g^2 \left(\Pi^{3 \gamma}(Q^2)-\Pi^{3 \gamma}(0)\right) \;.
\label{eq:dalpha_2}
\end{equation}
As for the purely electromagnetic current correlator, $\Pi^{3 \gamma}$ denotes the transverse part of the vacuum polarisation function.

Beyond the leading log
approximation, $\Delta \alpha_{\rm QED}^{\rm hvp}$ and $\Delta\alpha_2 ^{\rm hvp}$ become renormalisation scheme dependent. Additional hadronic
contributions to these
corrections at the scale of the W-mass and the Z-mass originate from chiral symmetry breaking. They have been
shown to be calculable in perturbation theory and to be at least two orders of magnitude smaller and thus negligible compared to the leading
contributions~\cite{Jegerlehner:1985gq}. Thus, having computed $\Delta \alpha_{\rm QED}$ before, all that is left to do to leading order is to
compute $\Delta \alpha_2$ as given in Eq.~(\ref{eq:dalpha_2}).

\subsection{Lattice calculation}
Since our ensembles feature mass-degenerate up and down quarks, $m_u = m_d$, light-quark disconnected contributions
cannot occur in $\Pi^{3\gamma}$ due to the isospin symmetry of the vacuum. Without those interference terms, single-flavour contributions to
Eq.~(\ref{eq:pi_zgamma}) in the continuum limit have the general structure $\Pi^{3\gamma, f} = < (V-A) V> $, where V and A denote vector and axial
vector currents,
respectively.
Since QCD conserves parity, no mixing between vector and axial vector currents occurs such that without
quark-disconnected contributions  we obtain
for up-type quarks twice the contribution of down-type quarks
\begin{equation}
 \Pi^{3 \gamma, \rm u}_{\mu \nu} = \frac{1}{6} \sum_f \langle\left(\overline{u}_f \gamma_{\mu} u_f\right)\left(\overline{u}_f \gamma_{\nu} u_f\right)\rangle
= 2 \Pi^{3 \gamma, \rm d}_{\mu \nu} \, .
\end{equation}

Combining this with $\Delta \alpha_{\rm QED}$, the leading-order hadronic contribution to the running of the weak mixing angle from the two light
flavours reads
\begin{equation}
 \Delta_{\rm ud}^{\rm hvp} (Q^2) = -\Delta^{\rm ud} \alpha_2^{\rm hvp}(Q^2) + \Delta^{\rm ud} \alpha_{\rm QED}^{\rm hvp}(Q^2) = \frac{1}{4} g^2 \Pi^{\rm uu}(Q^2) -
\frac{5}{9} e^2 \Pi^{\rm uu}(Q^2) \, .
\end{equation}
Neglecting disconnected contributions also for the heavy flavours, we have for the strange and the charm quark contributions
\begin{eqnarray}
 \Delta_{\rm s}^{\rm hvp} (Q^2) = -\Delta^{\rm s} \alpha_2^{\rm hvp}(Q^2) + \Delta^{\rm s} \alpha_{\rm QED}^{\rm hvp}(Q^2) & = \frac{1}{12} g^2 \Pi^{\rm ss}(Q^2) -
\frac{1}{9} e^2 \Pi^{\rm ss}(Q^2) \\ 
 \Delta_{\rm c}^{\rm hvp} (Q^2) = -\Delta^{\rm c} \alpha_2^{\rm hvp}(Q^2) + \Delta^{\rm c} \alpha_{\rm QED}^{\rm hvp}(Q^2) & = \frac{1}{6} g^2 \Pi^{\rm cc}(Q^2) -
\frac{4}{9} e^2 \Pi^{\rm cc}(Q^2) \; ,
\end{eqnarray}
respectively. 
Hence, the single flavour contributions are all proportional to the hadronic vacuum polarisation function but with different prefactors than for
$\Delta \alpha_{\rm QED}^{\rm hvp}$. In order to treat both contributions to $\Delta^{\rm hvp}\sin \theta_W$ consistently, we use in the light sector
the same redefinition of the vacuum polarisation function for $\Delta \alpha_2^{\rm hvp}$ as for $\Delta \alpha_{\rm QED}^{\rm hvp}$. 
Thus, in our lattice calculation we consider
\begin{equation}
\Delta^{\rm hvp}\sin^2 \theta_W (Q^2) =
  \Delta^{\rm hvp, ud} (Q^2 \cdot H^2 / H_\mathrm{phys}^2) +  \Delta^{\rm hvp, s} (Q^2) + \Delta^{\rm hvp, c} (Q^2)\,.
  \label{eq:delta_sin_theta_w_redefined}
\end{equation}

\subsection{Results}

\subsubsection{$\Delta \alpha_2^{\rm hvp}$}
As stated above, the leading hadronic contribution to the running of the weak mixing angle in the leading logarithmic approximation is obtained from
the difference of the corresponding contributions of the electromagnetic and the $SU(2)_L$ coupling constants, $\alpha_{\rm QED}$ and $\alpha_2$.
In contrast to $\Delta  \alpha_{\rm QED}^{\rm hvp}$, it is not straightforward to extract $\Delta  \alpha_2^{\rm hvp}$ from experimental $e^+ e^-
\rightarrow {\rm hadrons}$ data, since the data comprising the three lightest quarks would have to be separated either in up-type (u) and down-type
(d and s) quarks or assuming isospin symmetry in light and strange quark contributions. This problem has no unique solution, e.g.~final states
involving kaons could either originate directly from a strange quark current or from a gluon that could be radiated off light quarks.
Another possibility is to assume $SU(3)_f$ symmetry and thus only split the data into information attributed to the three lightest quarks and the
rest. The contributions from charm and heavier quarks can be computed in perturbation theory.

\begin{figure}[tb]
 \centering
\includegraphics[width=0.65\textwidth]{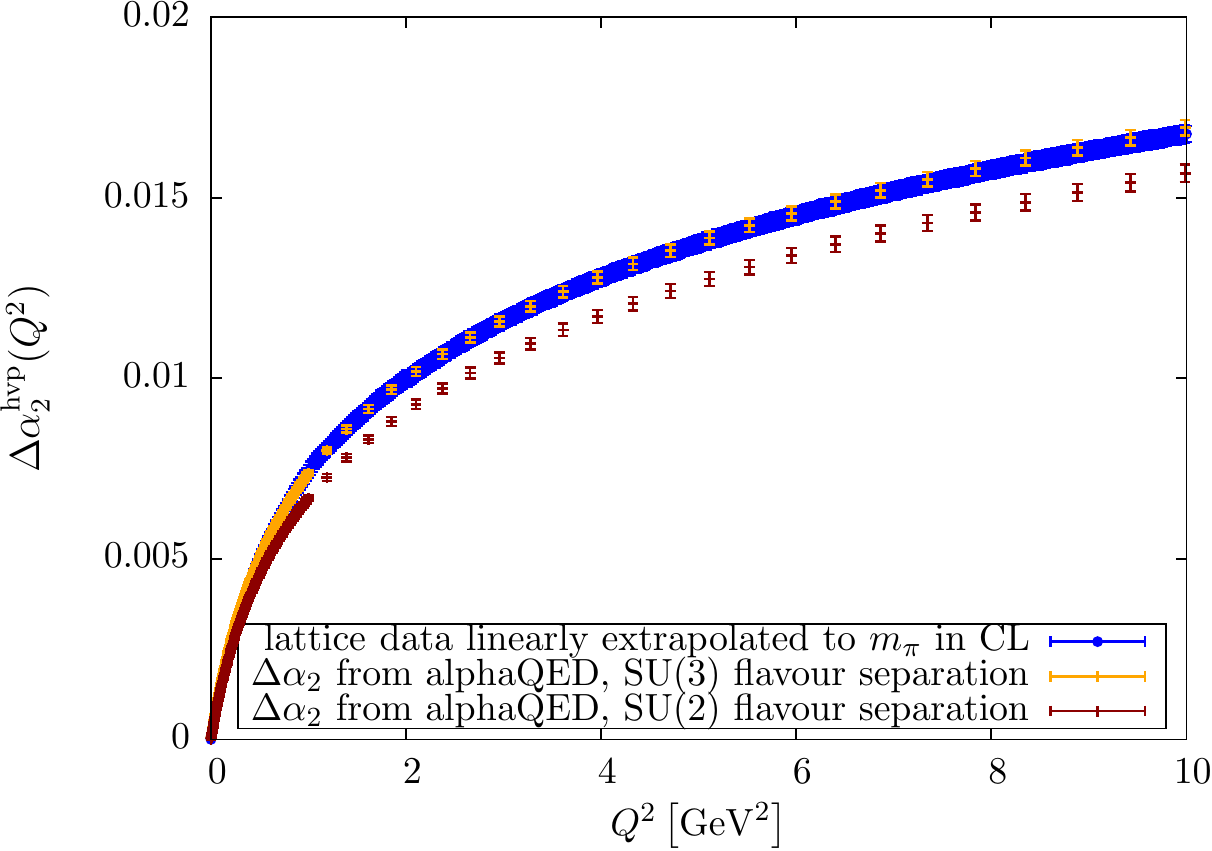}
\caption{$N_f=2+1+1$ contribution to $\Delta \alpha_2^{\rm hvp}$ compared to the data collected in~\cite{Jegerlehner:2012:Online} for all quarks
except the top.
The lattice data are extrapolated to the physical point and to the continuum limit (CL).}
\label{fig:dg_tot}
\end{figure}

Fig.~\ref{fig:dg_tot} shows our results after combined extrapolation to the physical point and to vanishing lattice spacing compared to the results
of~\cite{Jegerlehner:2011mw}. There, two ways of flavour separation have been implemented, one is assuming approximate $SU(3)_f$ and the other one
$SU(2)_f$ symmetry neglecting OZI violating terms. Our results clearly prefer the $SU(3)_f$ flavour separation and thus indicate that the
latter assumption is not tenable as has also been observed in~\cite{Francis:2013jfa} in a different context. In Fig.~\ref{fig:dg_tot} we have
multiplied the data from~\cite{Jegerlehner:2012:Online} with $\sin^2 \theta_W (M_Z) / \sin^2(\theta_0)$ to account for the different
reference values employed. As mentioned before $\sin^2(\theta_0)=0.23871(9)$ and the value used by Jegerlehner is $\sin^2\theta_W(M_Z) =
0.23153(16)$ which has been measured at LEP~\cite{ALEPH:2010aa}. The flavour
separation performed for the data set including very recent $e^+ e^-$ measurements is based on isospin symmetry relations~\cite{Jegerlehner:2015} and
the results are much closer to the ones based on $SU(3)_f$ flavour separation in Fig.~\ref{fig:dg_tot} than to the old $SU(2)_f$ curve. Thus, our
lattice results are also compatible with the newest phenomenological analysis based on an isospin $SU(2)_f$
flavour separation, however, not assuming flavour non-diagonal elements
to be small.

\subsubsection{$\Delta^{\rm hvp}\sin^2 \theta_W$}
Having determined the four-flavour contributions to $\Delta \alpha_{\rm QED}^{\rm hvp}$ and $\Delta \alpha_2^{\rm hvp}$, 
 it is straightforward to
obtain the leading-order hadronic vacuum polarisation contribution to the running of the weak mixing angle
\begin{equation}
 \Delta^{\rm hvp}\sin^2 \theta_W (Q^2)= \Delta \alpha_{\rm QED}^{\rm hvp}(Q^2) -\Delta \alpha_2^{\rm hvp}(Q^2) \; .
\end{equation}
This is the central observable
measured
in various low-energy experiments in order to gain hints on beyond the SM physics. In subsection~\ref{sec:res_theta} below, a selection of such
experiments
operating at momentum transfers investigated in this work will be listed.

\begin{figure}[tb]
 \centering
\includegraphics[width=0.73\textwidth]{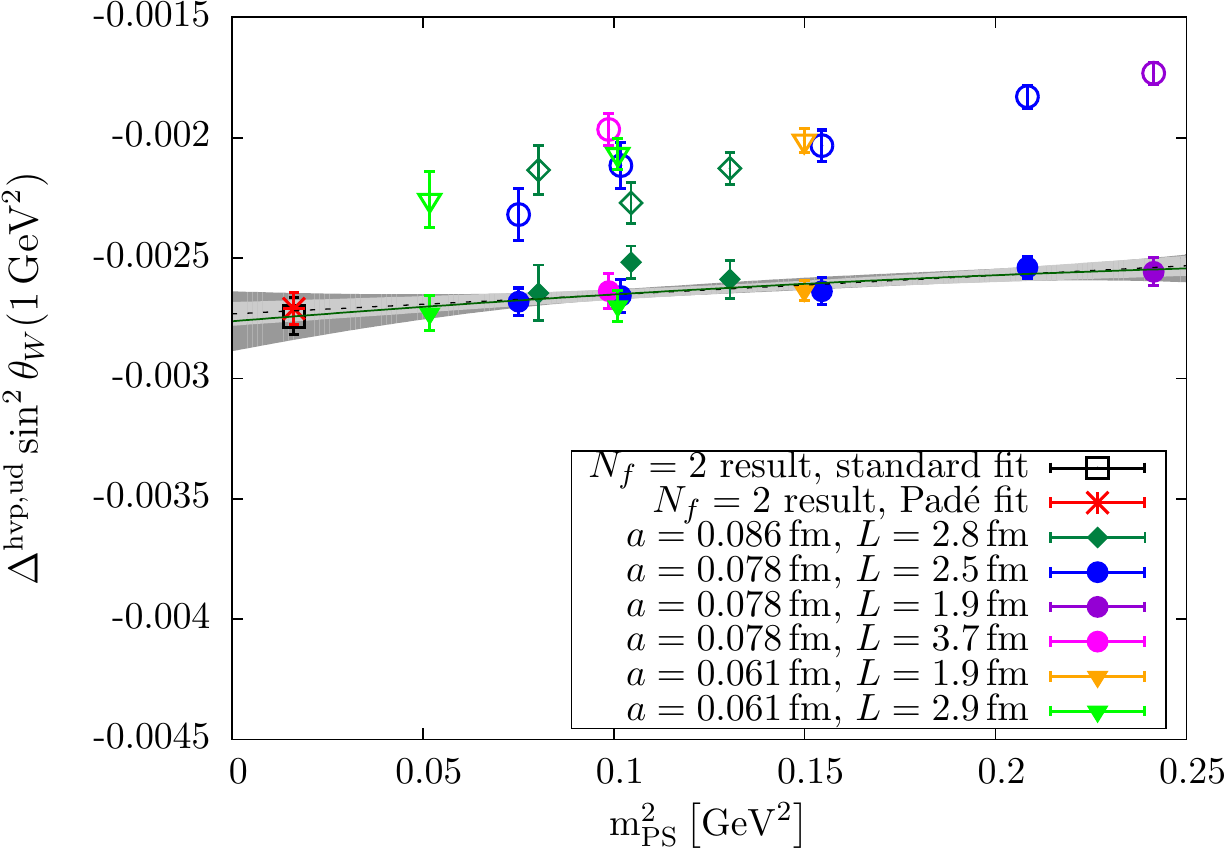}
\caption{Light-quark contribution to $\Delta^{\rm hvp} \sin^2\theta_W$ with
filled symbols representing points obtained with
Eq.~(\ref{eq:pi_redef}), open  symbols
refer to those obtained with  Eqs.~(\ref{eq:def_alpha}) and (\ref{eq:dalpha_2}). In particular, the
two-flavour results at the
physical point have been computed with the standard
definitions. The light grey errorband displays the uncertainty of the linear fit represented by the black dotted line wheras the dark grey errorband
belongs to the quadratic fit shown as the green solid line.}
\label{fig:theta_light}
\end{figure}

The physical results for the light-quark contribution for each momentum value can again be obtained from extrapolations in the squared pion mass as
shown in Fig.~\ref{fig:theta_light} for $Q^2=1\,{\rm GeV}^2$. In contrast to
the case of $\alpha_{\rm QED}$ depicted in the left panel of Fig.~\ref{fig:alpha_light_heavy}, for the weak mixing angle combining the redefinitions
according to Eq.~(\ref{eq:pi_redef}) of $\alpha_{\rm QED}$ and $\alpha_2$ leads to lower values than obtained with
the standard definitions. The common
feature of the leading-order hadronic contributions of both quantities is that the values procured with the redefinitions can be already
well-described by a simple linear extrapolation
in the squared pion mass to the physical point yielding a result which is compatible with those of the standard analysis as well as the one from
Pad\'e approximants on the ensemble of two dynamical quarks at the physical point. The results at the physical value of the pion mass are given in
table \ref{tab:theta_light}.

\begin{table}
\begin{center}
\begin{tabular}{|c|cc|}
 \hline
 $N_f=2+1+1$ extrapolated & $N_f=2$ standard & $N_f=2$ $[1,1]$ Pad\'e\\
 \hline
 \hline
 -0.002717(43) & -0.002742(78) & -0.002710(68) \\
 \hline
\end{tabular}
\end{center}
\caption{Comparison of results for $\Delta^{\mathrm{hvp, ud}}\sin^2\theta_W (1\,{\rm GeV}^2)$ at the physical point. The same analyses
as indicated
below
 table \ref{tab:alpha_light} have been performed.}
\label{tab:theta_light}
\end{table}

\begin{figure}[htb]
 \centering
\includegraphics[width=0.65\textwidth]{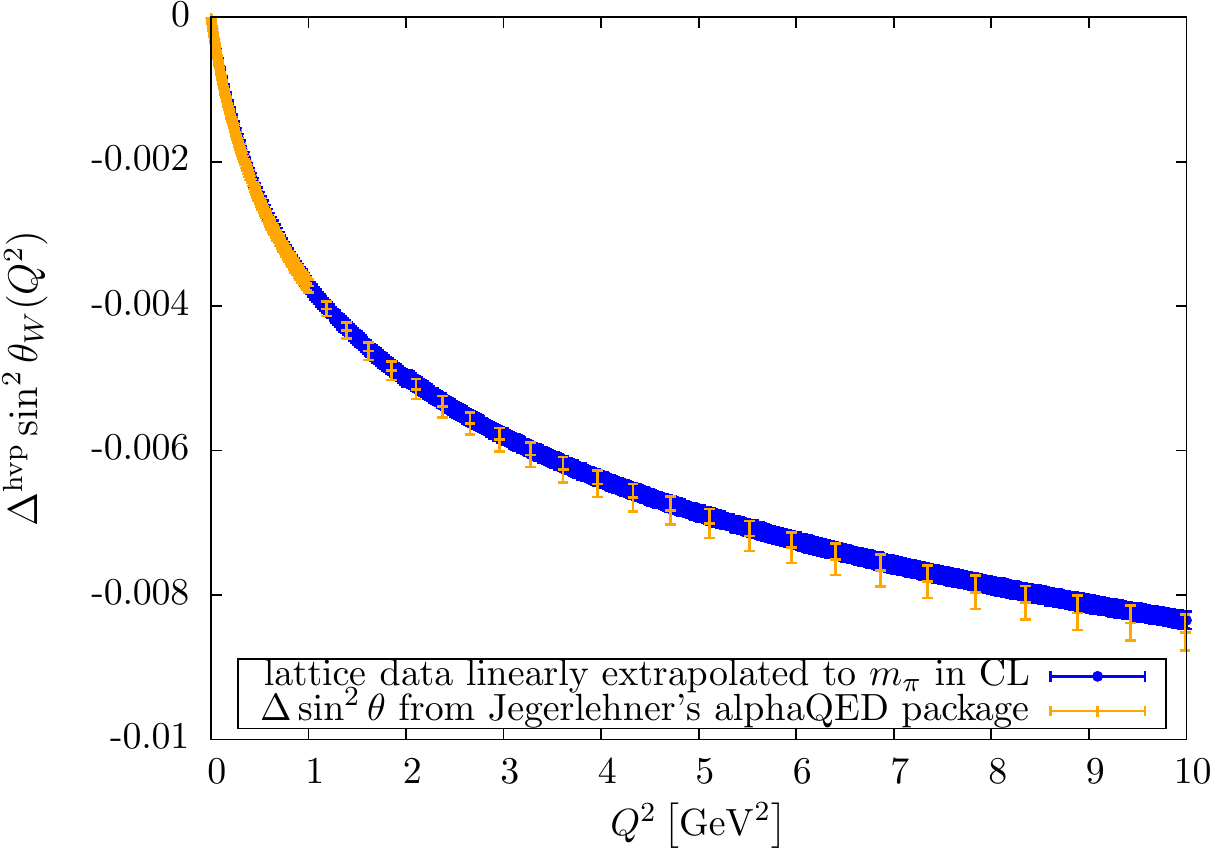}
\caption{$N_f=2+1+1$ contribution to the leading-order hadronic contribution $\Delta^{\rm hvp} \sin^2 \theta_W$ compared to the difference of the data
collected in~\cite{Jegerlehner:2012:Online}.The lattice data are extrapolated to the physical point and to the continuum limit (CL).}
\label{fig:theta_tot}
\end{figure}

When incorporating the heavy quarks, the chiral extrapolation is again combined with taking the continuum limit of the four-flavour result according
to
\begin{equation}
\Delta^{\rm hvp} \sin^2\theta_W (Q^2)(m_{\rm PS}, a)= A+ B~m_{\rm PS}^2 + C~a^2 \; .
\label{eq:theta_extrap}
\end{equation}
The results are shown in Fig.~\ref{fig:theta_tot}.
Complying with the indication from the previous subsection, we have
employed the results for $\Delta \alpha_2^{\rm hvp}$ obtained from $SU(3)_f$ flavour separation in
Fig.~\ref{fig:theta_tot} together with the factor needed to take the different reference values into account. Since we do not have information on the
correlation of the data in~\cite{Jegerlehner:2012:Online}, we have simply added the uncertainties of $\Delta \alpha_{\rm QED}^{\rm hvp}$ and $\Delta
\alpha_2^{\rm hvp}$ in quadrature  and may thus overestimate the errors of the phenomenological determination.

\subsubsection{Systematic uncertainties}
Since the systematic uncertainties stem from the same sources as for $\Delta \alpha_{\rm QED}^{\rm hvp}$ discussed before, the
relative errors are the same and only the absolute numbers differ due to the different prefactors of the renormalised vacuum polarisation function.
Naturally, also the plots all look very similar.
Therefore, we refrain from discussing the systematic effects separately and only summarise the general findings.

\begin{figure}[htb]
 \centering
\includegraphics[width=0.55\textwidth]{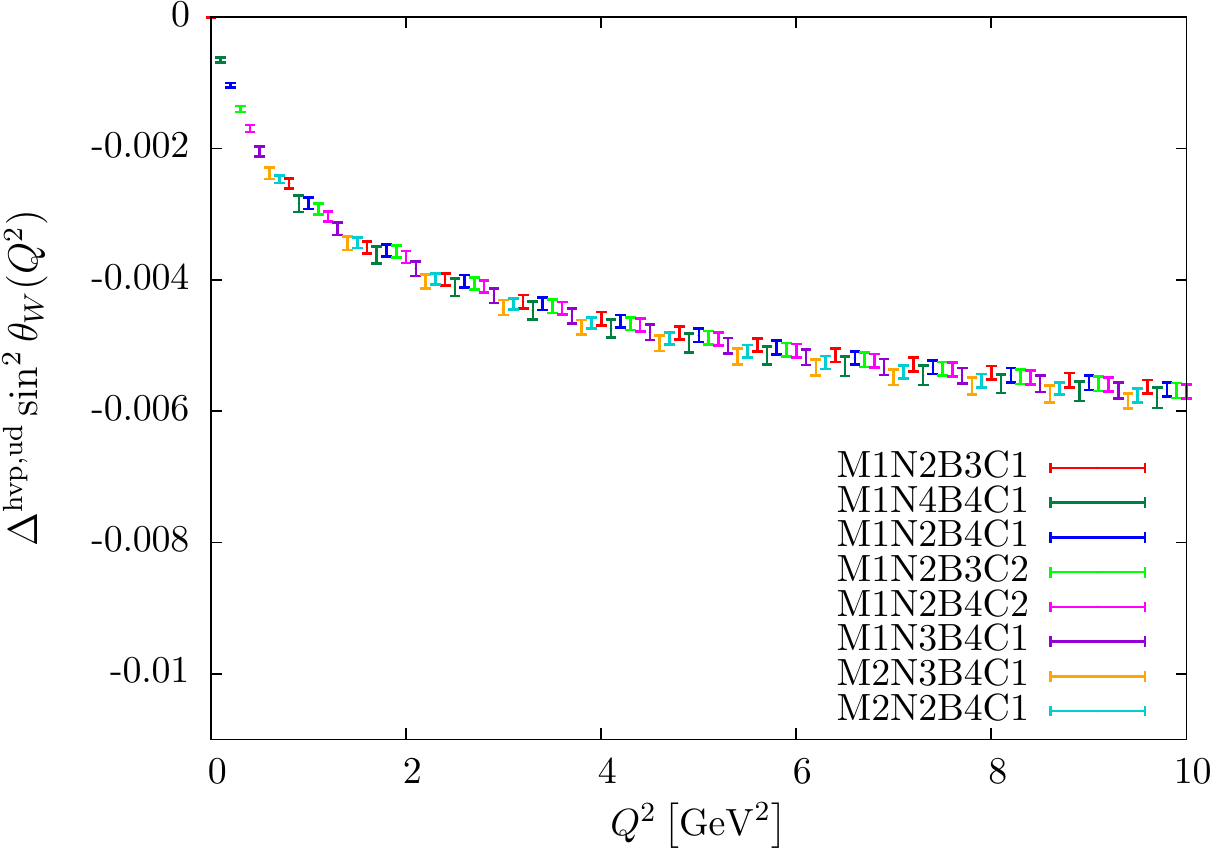}
\caption{Light quark contribution to $\Delta^{\rm hvp}\sin^2 \theta_W$ obtained from different fit functions. The
standard fit is M1N2B4C1.}
\label{fig:theta_light_mnbc}
\end{figure}

As before, due to the light quark contribution being an order of magnitude bigger than the contributions from the heavy quarks, we only need to take
systematic uncertainties of this part into account.
The dominant source of systematic errors is again the choice of the vacuum polarisation fit function as depicted in Fig.~\ref{fig:theta_light_mnbc}.

\begin{figure}[htb]
 \centering
\includegraphics[width=0.55\textwidth]{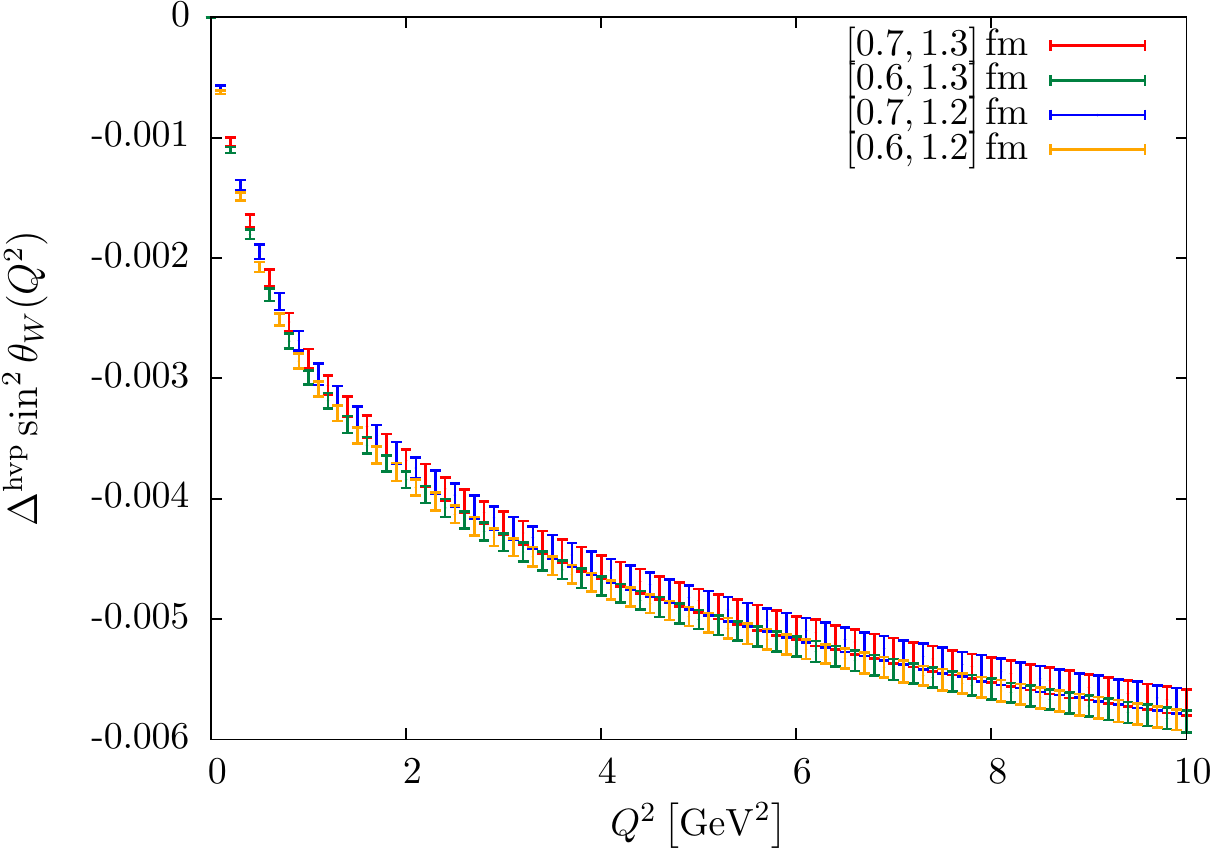}
\caption{Light quark contribution to $\Delta^{\rm hvp}\sin^2 \theta_W$ obtained with different fit ranges for the $\rho$ meson properties. The
standard fit range is $[0.7\, {\rm fm},1.2\,{\rm fm}]$.}
\label{fig:theta_light_fitranges}
\end{figure}

The only other relevant effect comes from the excited state contamination of the $\rho$-meson correlator and is shown in
Fig.~\ref{fig:theta_light_fitranges}. Finite volume effects and the choice of rather heavy pion masses in the chiral extrapolation seem to be
negligible in our calculation as outlined before.
The only unknown systematic effect is the heavy-flavour disconnected contributions which we have neglected here. 

\subsubsection{Final results for selected momentum values}
\label{sec:res_theta}
In table~\ref{tab:q2theta} we collect our results for $\Delta^{\rm hvp}\sin^2 \theta_W$ with statistical as well as systematic uncertainties for selected
momentum values. Experiments which have measured or will measure the weak mixing angle in the respective momentum region are also indicated. 

The
outcome of the E158 experiment at the SLAC linear accelerator was the first successful measurement of parity violation in electron-electron (M\o ller)
scattering~\cite{Anthony:2005pm}. The momentum transfer was $Q^2= 0.026\,{\rm GeV^2}$. The Qweak experiment conducted at JLAB in 2012
measured parity violation in electron-proton scattering at almost exactly the same momentum transfer~\cite{Armstrong:2014tna}. 
The data is still being
analysed. The predicted final uncertainty is about $5\%$ or $0.7\cdot 10^{-3}$ taking the central SM value. Another JLAB experiment
performed by the PVDIS collaboration determined the weak mixing angle from parity-violating deep inelastic scattering~\cite{Wang:2014bba,
Wang:2014guo} which effectively means electron-quark scattering at $Q^2=1.085\,{\rm GeV^2}$ and $Q^2=1.901\,{\rm GeV^2}$.
The envisioned successor of
the PVDIS experiment which also measures parity violation in electron-quark scattering is the SoLID spectrometer proposed at
JLAB~\cite{Chen:2014psa}. It can study about 20 kinematic points with $Q^2$ ranging from about $2\,{\rm GeV}^2$ to about $10\,{\rm GeV}^2$.
Its target accuracy is $6\cdot 10^{-4}$. Our results in table \ref{tab:q2theta} indicate that it will be essential to take the leading QCD corrections into
account in order to
deploy the whole potential of the experiment in the search for new physics beyond the SM.

\begin{table}
\centering
 \begin{tabular}{c | c c }
 \hline 
  & \vspace{-0.40cm} \\
 $Q^2$ [$\rm GeV^2$]& this work & experiment \\
 \hline \hline
 0.02 & $-0.158\,(05)\,(08)\cdot 10^{-3}$ & E158, Qweak \\
 1.00 & $-3.706\,(83)\,(127)\cdot 10^{-3}$ & PVDIS \\
 2.00 & $-5.021\,(96)\,(135)\cdot 10^{-3}$ & PVDIS \\
 3.00 & $-5.801\,(104)\,(135)\cdot 10^{-3}$& SoLID \\
 4.00 & $-6.398\,(102)\,(135)\cdot 10^{-3}$& SoLID \\
 6.00 & $-7.251\,(111)\,(136)\cdot 10^{-3}$& SoLID \\
 8.00 & $-7.867\,(112)\,(137)\cdot 10^{-3}$& SoLID \\
 10.0 & $-8.352\,(119)\,(138)\cdot 10^{-3}$& SoLID 
 \end{tabular}
\caption{ \label{tab:q2theta} We tabulate $\Delta^{\rm hvp}\sin^2 \theta_W$ for selected   
values of $Q^2$. The first                                             
error of the lattice results is statistical, the second systematic.  
Several low energy experiments sensitive to the respective momentum regions are indicated in the
last column.}
\end{table}

\section{Summary and Outlook}
\label{sec:conc}

Hadronic contributions to the running of electroweak parameters nowadays
constitute the major uncertainties of their values even at  
high energies thus also limiting the precision achievable in 
predictions for future high-energy colliders. Here we have considered the running 
of $\alpha_{\rm QED}$ and of the weak mixing angle which 
represents one of the most important parameters of the SM and 
provides a sensitive probe of new physics over a large energy range. 
 
Lattice QCD provides a most valuable tool to compute these hadronic contributions
from first principles alone. As we have demonstrated in this article, 
lattice QCD can be used to compute to a good precision the leading-order hadronic contribution
to the running of $\alpha_{\rm QED}$. 
In particular, we have carried out the 
first dynamical four-flavour calculation of the leading-order hadronic contribution to the 
running of the fine structure constant and the first lattice QCD 
calculation of the leading hadronic contribution to the shift of 
the weak mixing angle at energies between $0$ and $10\,{\rm GeV}^2$. 
In both cases the chiral as well as continuum extrapolations have been performed. 
A main effort has been undertaken to assess systematic uncertainties on a quantitative level.

For both quantities, agreement of our results with a phenomenological determination is 
observed with an even comparable statistical uncertainty.
However, we have found that the 
systematic effects of the calculation still exceed the statistical errors. 
The dominant systematic uncertainty has been found to be the choice of fit
function. Thus, methods which try to avoid fitting the vacuum polarisation are promising 
to reduce the overall uncertainty. A further improvement can be achieved
by increasing the 
statistical precision which would more strongly constrain the vacuum polarisation fit. 
Such improvements might be accomplished by the use of the
all-mode-averaging~\cite{Blum:2012uh} or the exact deflation~\cite{Saad:1984, Neff:2001zr} techniques.

In this article, we have provided further successful examples 
for the programme to determine hadronic contributions to electroweak
observables from lattice QCD. The steady progress in lattice QCD with ever increasing 
statistical accuracy and better understanding and control of systematic uncertainties
makes the lattice approach to compute these hadronic contributions very promising and 
gives hope that lattice results can directly be used for future low energy and 
collider experiments.

\section*{Acknowledgements}
We are most grateful to Fred Jegerlehner for very enlightening discussions. We thank the European Twisted Mass Collaboration (ETMC) for generating the
gauge field ensembles used for the calculations. This work has been supported in part by the DFG Corroborative
Research Center SFB/TR9.
G.P.~gratefully acknowledges the support of the German Academic National Foundation (Studienstiftung des deutschen Volkes e.V.) and of the
DFG-funded Graduate School GK 1504.

\bibliographystyle{JHEP}
\bibliography{alpha_and_theta}

\end{document}